%% file: main.tex
\newif\ifpreprint
\preprinttrue
\documentclass[sigconf,authorversion]{acmart}
\usepackage{multirow}
\usepackage{subcaption}
\usepackage{listings}
\usepackage{algpseudocode}
\usepackage[ruled,vlined,linesnumbered]{algorithm2e}
\usepackage[super]{nth}
\usepackage{colortbl} 
\usepackage{soul}
\usepackage{fnpct}
\usepackage{enumitem}
\usepackage{caption}
\usepackage[normalem]{ulem}

\usepackage{pifont}
\newcommand{\cmark}{\ding{51}}

\usepackage{bm}

\AtBeginDocument{%
  }

\copyrightyear{2026}
\acmYear{2026}
\setcopyright{cc}
\setcctype{by-nc-nd}
\acmConference[KDD 2026] {Proceedings of the 32nd ACM SIGKDD Conference on Knowledge Discovery and Data Mining V.1}{August 9--13, 2025}{Jeju Island, Republic of Korea.}
\acmBooktitle{Proceedings of the 32nd ACM SIGKDD Conference on Knowledge Discovery and Data Mining V.1 (KDD 2026), August 9--13, 2025, Jeju Island, Republic of Korea}
\acmISBN{979-8-4007-2258-5/2026/08}
\acmDOI{10.1145/3770854.3780313}

\acmSubmissionID{v1rtp1029}

\begin{document}
\title{
IRG: Modular Synthetic Relational Database Generation with Complex Relational Schemas
}

\author{Jiayu Li}
\orcid{0009-0003-2603-9421}
\affiliation{
\institution{University of Illinois, Urbana-Champaign}
\city{Champaign}
\state{IL}
\country{USA}
}\authornote{Email: jiayul11@illinois.edu. Work done while working at National University of Singapore and Betterdata AI.}

\author{Zilong Zhao}
\orcid{0000-0001-6549-3414}
\author{Milad Abdollahzadeh}
\orcid{0000-0003-4011-4670}
\affiliation{%
  \institution{Betterdata AI}
  \city{Singapore}
  \country{Singapore}
}




\author{Biplab Sikdar}
\orcid{0000-0002-0084-4647}
\author{Y.C. Tay}
\orcid{0000-0002-6280-2469}
\affiliation{
  \institution{National University of Singapore}
  \city{Singapore}
  \country{Singapore}
}
\renewcommand{\shortauthors}{Jiayu Li, Zilong Zhao, Milad Abdollahzadeh, Biplab Sikdar, \& Y.C. Tay}

\begin{abstract}
  \input{sections/0_abstract}

\end{abstract}



\received{7 February 2025}
\received[revised]{25 Jul 2025}
\received[accepted]{24 Nov 2025}


\maketitle

\section{Introduction}
\label{sec:intro}
\input{sections/1_introduction}

\vspace{-1em}
\section{Related Works}
\label{sec:related}
\input{sections/2_literature}
\section{Modeling RDB with CRS}
\label{sec:schema}
\input{sections/3_notations}

\section{IRG Data Generation}
\label{sec:irg}
\input{sections/4_irg}

\section{Experiments}
\label{sec:exp}
\input{sections/5_experiments}

\section{Conclusion}

\input{sections/6_conclusion}

\begin{acks}
We sincerely thank the Institute for Application of Learning Science and Educational Technology (ALSET) team from National University of Singapore, especially Mr. Kevin Hartman, for providing a motivation dataset for this work (private and therefore not presented in this paper).
We are also grateful to Dr. Uzair Javaid, CEO of Betterdata AI, for supporting this project. Additionally, we thank our current and former colleagues at Betterdata AI---Dr. Rishav Chourasia, Mr. Vikram Chundawat, and Dr. Bingyin Zhao---for engaging in valuable discussions that contributed to this work.
\end{acks}
\ifpreprint
\clearpage
\fi
\bibliographystyle{ACM-Reference-Format}
\ifpreprint
\bibliography{ref}
\fi

\ifpreprint
\clearpage
\fi
\appendix
\input{sections/7_appendix}

\end{document}

%% file: sections/0_abstract.tex
Relational databases (RDBs) are widely used by corporations and governments to store multiple related tables.
Their relational schemas pose unique challenges to synthetic data generation for privacy-preserving data sharing, e.g., for collaborative analytical and data mining tasks, as well as software testing at various scales.
Relational schemas typically include a set of primary and foreign key constraints to specify the intra-and inter-table entity relations, which also imply crucial intra-and inter-table data correlations in the RDBs.
Existing synthetic RDB generation approaches often focus on the relatively simple and basic parent-child relations,
failing to address the ubiquitous real-world complexities in relational schemas in key constraints like composite keys, intra-table correlations like sequential correlation, and inter-table data correlations like indirectly connected tables.
In this paper, we introduce \textbf{i}ncremental \textbf{r}elational \textbf{g}enerator (IRG), a modular framework designed to handle these real-world challenges. In IRG, each table is generated by learning context from a depth-first traversal of relational connections to capture indirect inter-table relationships and 
constructs different parts of a table through several classical generative and predictive modules to preserve complex key constraints and data correlations.
Compared to 3 prior art algorithms across 10 real-world RDB datasets, IRG successfully handles the relational schemas and captures critical data relationships for all datasets while prior works are incapable of.
The generated synthetic data also demonstrates better fidelity and utility than prior works, implying its higher potential as a replacement for the basis of analytical tasks and data mining applications.
Code is available at: \url{https://github.com/li-jiayu-ljy/irg}.

%% file: sections/1_introduction.tex
Relational databases (RDBs) are among the major data structures in which most modern-day corporations, governments, and organizations store their collected data when they come in multiple related tables~\cite{rdb-data}. However, the utility of real data is often limited. For instance, the General Data Protection Regulation (GDPR)~\cite{gdpr} imposes strict limitations on sharing user-collected data with external collaborators due to privacy concerns. A straightforward solution to this challenge is the use of synthetic data. Since synthetic entities are not directly related to real-world individuals, they can be shared more freely~\cite{pdpc,sec-syn}. Moreover, synthetic data can strengthen knowledge discovery pipelines by empowering deep-learning models to solve challenging problems in database mining~\cite{syn-kdd}. Additionally, for purposes such as software testing—whether for sanity checks or stress tests—a scaled-down or scaled-up version of the dataset is often more practical than the original. Synthetic data, with its inherent flexibility in scaling, effectively addresses these challenges.

Compared to the generation of other data formats (e.g., tabular, images), RDBs pose unique challenges due to the existence of relational schemas that connect multiple tables based on certain rules, which also incur intra-and inter-table correlations. While existing synthetic RDB generators successfully capture the basic form of relational schemas—foreign key (FK) constraints between parent and child tables, they overlook the prevalent presence of complex relational schemas (CRSs)~\cite{sdv,rctgan,clavaddpm} (details see Sec.~\ref{sec:related} and \ref{ssec:crs_dataset}).

\begin{figure*}[t]
    \begin{minipage}{0.62\linewidth}
            \centering
    \captionof{table}{Summary of capabilities of existing models vs. IRG. \cmark means that the model can directly handle the case, \cmark${}^*$ indicates that additional simple feature engineering on the dataset is required to enable the model to handle the cases. 
    }
    \label{tab:comparison}
    \vspace{-1.3em}
    \setlength{\tabcolsep}{2pt}
    \resizebox{0.95\linewidth}{!}{
\input{tables/comparison}
    }
    \end{minipage}
    \hfill
    \begin{minipage}{0.37\linewidth}
            \centering
    \includegraphics[width=\linewidth]{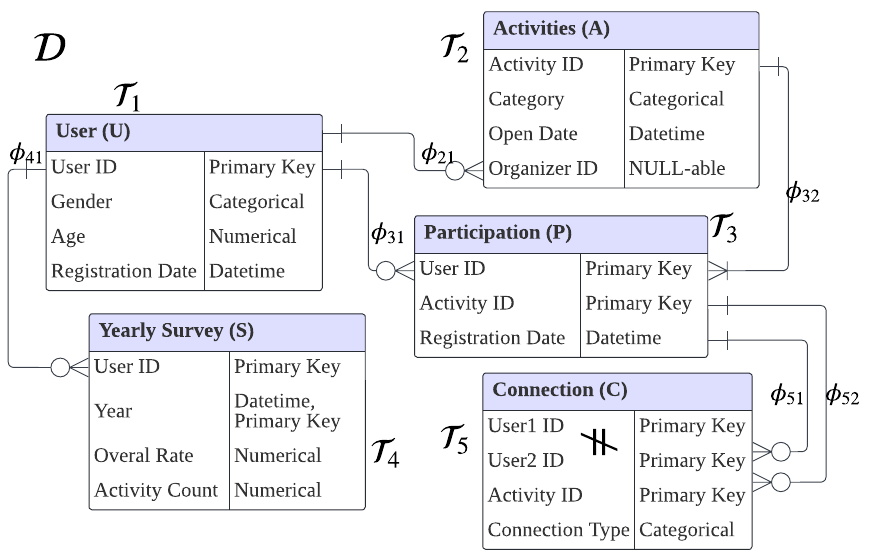}
    \vspace{-2.5em}
    \caption{Relational schema of an example RDB. 
    Multiple ``Primary Key''s in a table means composite PK.
    Table C contains two overlapping composite FKs with non-self-match constraint.
    }
    \label{fig:example-schema}
    \end{minipage}
    \vspace{-2em}
\end{figure*}

The relational schema of an RDB consists of names and domains of the relations and their attributes and relevant constraints, such as primary key (PK) and FK constraints~\cite{rdb-intro}. The names and domains are typically easy to capture by a tabular data processor, but the constraints are much trickier. Prior RDB generation frameworks often capture basic relational constraints, including an ID attribute as PK, and FK with a parent table and a child table, but fail to consider complex ones~\cite{sdv,rctgan,clavaddpm}, including but not restricted to:

\begin{itemize}[topsep=4pt, partopsep=0pt, itemsep=0pt, parsep=0pt,leftmargin=*]
    \item FKs with NULL (e.g., \texttt{transaction}s online do not have a \texttt{store} record so the \texttt{store ID} is NULL);
    \item Two or more FKs with the same parent and child tables (e.g., \texttt{sender} and \texttt{receiver} of \texttt{message}s):\begin{itemize}
        \item Multiple FKs with non-self-match constraint (e.g., \texttt{user}s cannot be  \texttt{friend}s with themselves);
    \end{itemize}
    \item Composite PKs (i.e., PKs involving multiple attributes):\begin{itemize}
        \item Two or more FKs forming a PK (e.g., for a \texttt{course selection} table, \texttt{course ID} and \texttt{student ID} pair should be unique);
        \item One or more FKs with a local serial ID forming a PK (e.g.,  \texttt{customer ID} and \texttt{transaction} time (i.e., local serial ID) for a \texttt{transaction}s' table);
    \end{itemize}
    \item Composite FKs (i.e., FKs involving multiple attributes, typically by referencing a table whose PK is composite):\begin{itemize}
        \item Two or more composite FKs having overlapping attributes (e.g., for \texttt{shot} table in a football RDB, FKs (\texttt{gameID}, \texttt{shooterID}) and (\texttt{gameID}, \texttt{assisterID}) have an overlapping attribute);
        
    \end{itemize}
\end{itemize}

Note that satisfaction of PK and FK constraints is necessary for the validity of synthetic RDB, so the negligence of and hence incapability in these CRSs render existing RDB generators essentially incapable of handling complex cases. Unfortunately, these CRSs are prevalent in real-world datasets, implying a demand for a new framework to handle CRSs.
\ifpreprint

\fi
The complexity of an RDB also comes from the number and type of tables and connections. Together with the CRSs above, the complexity of the RDB schemas also induce a complexity in data correlations, both intra-and inter-table. These complex correlations include but are not restricted to:

\begin{itemize}[topsep=4pt, partopsep=0pt, itemsep=0pt, parsep=0pt,leftmargin=*]
    \item Sequential correlation between rows in a child table with the same parent (e.g., \texttt{transaction}s of a \texttt{customer});
    \item Correlation between table values and shapes (sizes as in number of rows), as the table size should be similar to the real one while being related to the parent tables (e.g., the number of \texttt{transaction}s in total and per \texttt{customer} respectively);
    \item Correlation between step-parent (i.e., another parent of a sibling table (table sharing the same parent)) and step-children, which may never be introduced in frameworks that join tables always following topological order, which is the case of existing works.
\end{itemize}

To solve all the above-mentioned problems, we propose a novel RDB generation framework: \textbf{i}ncremental \textbf{r}elational \textbf{g}enerator (IRG). 
IRG constructs a more complete context from related tables of an RDB with CRS by a depth-first traversal of related tables, 
and decomposes the generation task for each table into 5 major components (modules) to rigorously address different CRS constraints:
\begin{enumerate}[topsep=4pt, partopsep=0pt, itemsep=0pt, parsep=0pt,leftmargin=*]
    \item Degree (number of child rows corresponding to the same parent row) generation with a controlled sum of degrees for all FKs;
    \item NULL-indicator generation of FKs for all NULL-able FKs\ifpreprint (FKs containing NULL values)\fi;
    \item Conditional tabular data generation, which is central to many prior works~\cite{rctgan,realtabformer,clavaddpm}, is used here to generate \textit{aggregated} information (details in Sec.~\ref{ssec:aggregated}), grouped by the first FK;
    \item Sequential data generation with static conditions (e.g., generating \texttt{transaction}s over time based on \texttt{customer} profiles);
    \item Obtaining other FK attributes' values by matching to their parent tables that conform to all relational schema constraints.
\end{enumerate}

\begin{figure}
    \centering
    \vspace{0.5em}
    \includegraphics[width=0.6\linewidth,trim={0 0.1em 0 0.4em},clip]{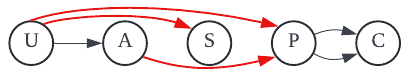}
    \vspace{-1em}
    \caption{Directed acyclic graph (DAG) corresponding to the \textit{example RDB} (Fig.~\ref{fig:example-schema}). Red edges illustrate the step-parent-child relation between A and S via A$\rightarrow$P$\rightarrow$U$\rightarrow$S.}
    \label{fig:eg-dag}
    \vspace{-0.5em}
\end{figure}
To summarize, our contributions are as follows: 
\begin{itemize}[topsep=4pt, partopsep=0pt, itemsep=0pt, parsep=0pt,leftmargin=*]
    \item We formalize RDB modeling with CRSs for data generation.
    \item 
    We propose a novel modular framework that integrates standard machine learning (ML) tasks into a single system for RDB generation. This framework is capable of understanding CRSs, and its components can be replaced with the latest methods.
    \item Analysis of 10 real-world RDBs shows that IRG is applicable to all 10 datasets, while other algorithms are only applicable to at most 1 dataset. We empirically demonstrate that our framework achieves a 100\% valid relational schema in synthetic data with high-quality results, whereas previous frameworks fail to do so.
\end{itemize}

%% file: tables/comparison.tex
    \begin{tabular}{ll|ccc|>{\columncolor[gray]{0.9}}c}
        \toprule
        & & SDV & RCTGAN & ClavaDDPM & IRG \\
        & & \cite{sdv} & \cite{rctgan} & \cite{clavaddpm} & (ours) \\
        \midrule
        \multirow{7}{*}{Key constraints} 
        & FK w/ NULL & \cmark & \cmark${}^*$ & \cmark${}^*$ & \cmark \\
        & $\ge2$ FKs w/ same parent and child tables & \cmark & \cmark &  & \cmark\\
        & $\ge2$ FKs w/ non-self-match constraint & & & & \cmark \\
        & PK/FK w/ $\ge2$ attributes & \cmark${}^*$ & \cmark${}^*$ & \cmark${}^*$ & \cmark  \\
        & $\ge2$ FKs form a PK & & & & \cmark \\
        & $\ge1$ FK \& a local ID as PK & & & & \cmark \\
        & $\ge2$ composite FKs w/ overlapping attributes & & & & \cmark \\
        \hline
        \multirow{3}{*}{Data correlations} & Sequential relation between rows with same parent & & & & \cmark\\
        & Table shapes integrity & & & & \cmark \\
        & Relation between step-parent and step-child & & & & \cmark \\
        \hline
        \multirow{2}{*}{Other use cases} & Scalable up to large number of tables & & & \cmark & \cmark\\
        & Flexibly configurable size per table during generation & & & & \cmark \\
        \bottomrule
    \end{tabular}

%% file: sections/2_literature.tex
There are two major paradigms of RDB data generation: \textbf{i) conditional generation:} construct conditions from parent tables to generate values in the child tables, so as to address the relational schema through parent-child relations; \textbf{ii) constrained generation:} regularize the relational schema constraints by loss functions.

For the first paradigm, REaLTabFormer~\cite{realtabformer} leverages conditional tabular generation, where rows from parent tables are treated as conditions and rows from child tables as values, using an auto-regressive transformer~\cite{transformer}. However, it focuses only on the parent-child relationship, without a complete solution for RDB generation.
PrivBench~\cite{spn} uses a sum-product network~\cite{basespn} for RDB generation, though it provides a complete solution, also essentially focuses on the parent-child relationship only, leaving it inapplicable to databases with slightly more complex schemas.
%
SDV~\cite{sdv,sdv-new} represents another framework within this paradigm, based on pure statistical methods.
Recent works have also introduced deep generative models to RDB generation. For instance, RCTGAN~\cite{rctgan} introduces Generative Adversarial Networks (GANs)~\cite{gan} and ClavaDDPM~\cite{clavaddpm} introduces diffusion models~\cite{diffusion}, to capture parent-child relations and extend to complete RDBs. 
For tables with two parents in a many-to-many relation, graph-based modeling is found to be an effective variant of conditional generation~\cite{m2m,rgcld}.
IRG also adopts conditional generation to synthesize simple parent-child relations. However, none of the works within the first paradigm thoroughly address CRSs (e.g., composite keys), limiting their ability to handle many real-world datasets.
Table~\ref{tab:comparison} summarizes the capabilities and limitations of some works compared to ours.

ITS-GAN~\cite{faketables}, FK-GAN~\cite{fk-gan}, and FakeDB~\cite{fakedb} are RDB generation models in the second paradigm. These are GANs with regularized loss to enforce schema constraints by functional dependency and first-order logic. Although schema constraints can usually be formulated for these models, they do not guarantee 100\% correctness. Therefore, they are not used as the building block of IRG.


\begin{table*}[h!t]
    \centering
    \caption{Presence of CRSs in real-world datasets and applicability of RDB generation methods. 
    \cmark indicates scenario existence or model applicability (ensuring all schema constraints, including PKs and FKs).
    ${}^*$ denotes applicability with simple feature engineering. ? indicates applicability but with overlooked data correlations (marked ``(?)''), contributing half to ``Count.''
    }
    \label{tab:complex-data}
    \vspace{-1em}
    \setlength{\tabcolsep}{1pt}
    \resizebox{\linewidth}{!}{
    \input{tables/datasets}
    }
    \vspace{-1em}
\end{table*}

\input{algos/multifold-theorem}

While the algorithms mentioned above primarily concentrate on framework design for general RDB generation, there are also works that target specific challenges. DScaler~\cite{dscaler} applies graph scaling techniques to address RDB scaling. \cite{priv-query-aware} integrates differential privacy into RDB generation to mitigate privacy concerns. ezGen~\cite{query-aware} decomposes complex queries into cardinality constraints to address query-aware RDB generation. However, since this paper focuses on RDB generation for general purposes, these specialized works are not included as comparative baselines in our study.

%% file: tables/datasets.tex
\begin{tabular}{l|cccccccccc|>{\columncolor[gray]{0.9}}c}
    \toprule
    Dataset & stack~\cite{relbench} & trial~\cite{relbench} & F1~\cite{relbench} & event~\cite{relbench} & avito~\cite{relbench} & bike~\cite{bike} & football~\cite{football} & social~\cite{social} & SMM~\cite{smm} & BEC~\cite{olist} & Count \\
    \midrule
    FK w/ NULL${}^*$ & \cmark & & & & \cmark & & \cmark & & \cmark & & 4 \\
    $\ge2$ FKs w/ same parent \& child tables & \cmark & & & \cmark & & & \cmark & \cmark & & & 4 \\
    $\ge2$ FKs w/ non-self-match constraint & \cmark & & & \cmark & & & \cmark & \cmark & & & 4 \\
    PK w/ $\ge2$ attributes & & \cmark & \cmark & \cmark & \cmark & \cmark & \cmark & \cmark & \cmark & \cmark & 9 \\
    FK w/ $\ge2$ attributes${}^*$ & & \cmark & & & & \cmark & \cmark & & \cmark & & 4\\
    $\ge2$ FKs form a PK & & \cmark & \cmark & \cmark & \cmark & \cmark & \cmark & \cmark & \cmark & & 8\\
    $\ge1$ FK \& a local ID as PK & & & \cmark & \cmark & \cmark & & & & & \cmark & 4 \\
    $\ge2$ FKs w/ overlapping attributes & & \cmark & & & & \cmark & \cmark & & \cmark & & 4\\
    Sequential relation within a table (?) & \cmark & \cmark & \cmark & \cmark & \cmark & \cmark & \cmark & \cmark & \cmark & \cmark & 10\\
    Existence of multi-fold connected trail (?) & \cmark & \cmark & \cmark & \cmark & \cmark & \cmark & \cmark & \cmark & \cmark & \cmark & 10 \\
    Self-reference${}^*$ & \cmark & & & & & \cmark & & & & & 2\\
    \midrule
    SDV applicability & ? & & & & & & & & & & 0.5 \\
    RCTGAN applicability & ? & & & & & & & & & & 0.5 \\
    ClavaDDPM applicability & & & & & & & & & & & 0 \\
    IRG (ours) applicability & \cmark${}^*$ & \cmark & \cmark & \cmark & \cmark & \cmark${}^*$ & \cmark & \cmark & \cmark & \cmark & \textbf{10} \\
    \bottomrule
\end{tabular}

%% file: algos/multifold-theorem.tex
\begin{algorithm}[t]
    \small
    \caption{Pseudocode of algorithm $\mathcal{A}$ in Thm.~\ref{thm:multi-fold}}\label{alg:multi-fold}
    \DontPrintSemicolon
    \KwIn{
    $\mathcal{D}=\{\mathcal{T}_1,\mathcal{T}_2,\dots,\mathcal{T}_M\}$,
    $\varphi=\{\Phi_1,\Phi_2,\dots,\Phi_M\}$\;
    }
    Program without involving $\varphi$ (e.g., initialization of global variables) \;
    \ForEach{$l\in\{1,2,\dots,L\}$}{
    Program without involving $\varphi$\;
    \For{$\phi$ in $\sigma_l$}{
    \Comment{$\sigma_l$ is a top-down/bottom-up topological order of $\varphi$}\;
    $f_l(\phi)$
    }
    \vspace{-0.3em}
    }
    Program without involving $\varphi$\;
\end{algorithm}

%% file: sections/3_notations.tex
In this section, we first define and formalize the relational schema. Next, we provide an empirical analysis of the presence of CRSs in 10 RDBs. Finally, we introduce a table relation based on a topological order and context construction for RDBs, which are essential for capturing the CRSs during both training and generation.

\begin{figure}[t]
    \vspace{-1em}
    \centering
    \begin{subfigure}[b]{0.29\linewidth}
        \centering
        \includegraphics{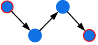}
        \vspace{-0.4em}
        \caption{\textit{1-fold connected}.}
    \end{subfigure}%
    ~ 
    \begin{subfigure}[b]{0.29\linewidth}
        \centering
        \includegraphics{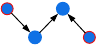}
        \vspace{-0.4em}
        \caption{\textit{2-fold connected}.}
    \end{subfigure}
    ~
    \begin{subfigure}[b]{0.29\linewidth}
        \centering
        \includegraphics{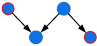}
        \vspace{-0.4em}
        \caption{\textit{3-fold connected}.}
    \end{subfigure}
    \vspace{-1.2em}
    \caption{Example of $n$-fold connected trails with 3 edges (two ends are highlighted in red circles).}
    \label{fig:nfold}
    \vspace{-0.5em}
\end{figure}

\subsection{Preliminaries and Schema Notations}
\label{sec:schema:notation}
\textbf{Relational Database Schema~\cite{rdb-intro}.} A relational database (RDB) $\mathcal{D}$ consists of $M$ tables (i.e., relations), $\{\mathcal{T}_1,\mathcal{T}_2,\dots,\mathcal{T}_M\}$. Each table $\mathcal{T}_i$ is a collection of $n_i$ rows (i.e., samples), where each row is described by a set of attributes (i.e., columns) $\mathcal{C}_i$ of size $|\mathcal{C}_i|$.
A \textbf{\textit{primary key}} (PK) is a column (or set of columns) $\mathcal{P}_i\subseteq\mathcal{C}_i$ in each table that uniquely identifies a row and can not be NULL (i.e., must always have a value).
Many tables have $|\mathcal{P}_i|=1$, where the PK is an ID, but PKs with multiple attributes, i.e., $|\mathcal{P}_i|>1$, are permissible, typically induced by relations with other tables\footnote{Note that it is also possible for a table to have no PK, in which case we denote $\mathcal{P}_i=\emptyset$ for simplicity.}.
For example, in Fig.~\ref{fig:example-schema}, (\texttt{User ID, Activity ID}) pair is PK of table P ($\mathcal{T}_3$).

A \textbf{\textit{foreign key}} (FK) in table $\mathcal{T}_i$ is a column in this table that refers to the PK in another table (e.g., table $\mathcal{T}_j$). FK creates a relation between tables $\mathcal{T}_i$ and $\mathcal{T}_j$, which allows us to relate the records across tables in our RDB. 
In this case, for every row $\mathbf{r}_i\in\mathcal{T}_i$, there exists exactly one row $\mathbf{r}_j\in\mathcal{T}_j$ whose PK value equals $\mathbf{r}_i$'s FK value. Here, $\mathcal{T}_i$ and $\mathcal{T}_j$ are also called the child and parent tables, and $\mathbf{r}_i$ is called a child of $\mathbf{r}_j$.
For example, in Fig.~\ref{fig:example-schema}, \texttt{User ID} is FK of table S ($\mathcal{T}_4$), which refers to the PK of table U ($\mathcal{T}_1$), so $\mathcal{T}_4$ is a child of $\mathcal{T}_1$.
%
%
Each table can have multiple FKs. 
Let $\Phi_i$ represent the set of FKs in $\mathcal{T}_i$, where $|\Phi_i|=K_i$. For the $k$-th FK $\phi_{ik}\in\Phi_i$, we denote the parent table as $\mathcal{T}_{F_{ik}}$ where $F_{ik}\ne i$ is the parent table's index. 
The FK attributes of $\phi_{ik}$ is the set $\mathcal{K}_{ik}\subseteq\mathcal{C}_i$ 
and the corresponding attributes from parent are $\mathcal{F}_{ik}\subseteq\mathcal{C}_{F_{ik}}$, and usually $\mathcal{F}_{ik}=\mathcal{P}_{F_{ik}}$. A valid FK must satisfy $|\mathcal{K}_{ik}|=|\mathcal{F}_{ik}|\ge1$, and when $|\mathcal{K}_{ik}|>1$, $\phi_{ik}$ is a composite FK. 
\ifpreprint
All notations introduced above are summarized in Table~\ref{tab:notations} in App.~\ref{app:notation}. To aid illustration, we use an \textit{example RDB}, shown in Fig.~\ref{fig:example-schema}, with further details and illustrations in App.~\ref{app:example:config}.
\fi

\noindent\textbf{Graph Modeling of RDB.} An RDB can be represented as a \textit{directed acyclic graph} (DAG) using the tables and FKs, where tables form vertices and FKs form edges $\mathcal{E}$ from parent to child tables:
\vspace{-0.5em}
\begin{equation}
    \mathcal{E} = \{(\mathcal{T}_{F_{ik}} \rightarrow \mathcal{T}_i)\text{ via }\phi_{ik} \, | \phi_{ik}\in\bigcup_{j=1}^M\Phi_j\}
\vspace{-0.5em}
\end{equation}
For example, the \textit{example RDB} is modeled as the DAG shown in Fig.~\ref{fig:eg-dag}. In this paper, we only consider RDBs without cyclic dependencies, including self-dependencies. Any RDB with cyclic dependency can be transformed acyclic by inserting auxiliary tables~\cite{dag}.

\noindent\textbf{RDB Generation Problem.} Given a well-defined RDB $\mathcal{D}$, the objective is to obtain a synthetic counterpart $\mathcal{D}'$. $\mathcal{D}'$ should have the same schema as $\mathcal{D}$, including tables' and attributes' names, and PK and FK constraints. All schema constraints, e.g., PKs and FKs, should be satisfied. $\mathcal{D}'$ is expected to be similar to $\mathcal{D}$ in terms of intra-and inter-table data distribution and correlations.

\subsection{Complex Relational Schemas Formalization}
\label{ssec:crs_dataset}
\ifpreprint
Most CRSs featured by complex key constraints can be formalized using the notations defined above. The exact expressions are provided in Table~\ref{tab:complex-notation} in App.~\ref{app:notation}. Most of them are also present in the \textit{example RDB}, with details summarized in Table~\ref{tab:eg-complex-only} in App.~\ref{app:example:config}.

\fi
Recall from Sec.~\ref{sec:intro} that a common inter-table data correlation, also a CRS scenario, that is often overlooked in prior works, is step-parent-child relations,
which can be generalized to multi-fold connected tables defined as follows, also illustrated in Fig.~\ref{fig:nfold}.
\begin{definition}
    Tables $\mathcal{T}_i$ and $\mathcal{T}_j$ are called $\bm{n}$\textbf{\textit{-fold connected}} in RDB $\mathcal{D}$, if they are connected by the DAG of $\mathcal{D}$, and there exists a \textit{trail} (i.e., a walk where no edge is repeated) from $\mathcal{T}_i$ to $\mathcal{T}_j$, 
    experiencing $n-1$ changes of direction on the trail\footnote{Two tables can be both $n_1$-fold and $n_2$-fold connected if multiple distinct trails exist.}.
\end{definition}
\begin{definition}
    Two tables are \textit{\textbf{multi-fold connected}} if they are $n$-fold connected with $n\ge3$.
\end{definition}

Multi-fold connected tables
are indirect but significant in real-world scenarios.
In the \textit{example RDB}, A and S are multi-fold connected ($n=3$) through A$\rightarrow$P$\rightarrow$U$\rightarrow$S (see Fig.~\ref{fig:eg-dag}). 
This connection from A to S matters because the kind of activities (A) a user participates in are highly relevant to the outcomes of the survey (S).

\newcommand{\thmmultifold}{

    Let $\mathcal{A}$ be an algorithm to learn the data distribution and relation of RDB $\mathcal{D}$. If all processes involving FKs can be described as a sequence of $L$ functions, $f_1,f_2,\dots,f_L$ that takes in an FK only and the functions are called according to some topological order of FKs, $\sigma_1,\sigma_2,\dots,\sigma_L$, satisfying the following constraints (see Algo.~\ref{alg:multi-fold}):
    \begin{enumerate} [topsep=0pt, partopsep=0pt, itemsep=0pt, parsep=0pt,leftmargin=*]
        \item \textbf{Only FK:} $f_l$ may access the parent and child table of FK and intermediate global variables, but not information of other FKs;
        \item \textbf{Topological Order:} $\sigma_l$ can be either topologically forward (top-down) or backward (bottom-up), where the order of the FKs (edges) refers to the order of the involved tables (vertices);
        \item \textbf{Invariant of Order:} For a complete iteration of each $f_l$, the result, including the modification on global intermediate variables, is invariant\footnote{We refer ``invariant'' in terms of content, e.g., changes in attribute order of a table are not considered ``variant'', but additional attributes or rows are.} for any value $\sigma_l$ as long as the direction (forward vs. backward) is the same.
    \end{enumerate}
    Then, there exist two connected tables $\mathcal{T}_i,\mathcal{T}_j$ on the DAG of some $\mathcal{D}$ whose joint distribution (i.e., $p(\mathcal{T}_i,\mathcal{T}_j)$) cannot be captured by $\mathcal{A}$. 
}
\begin{theorem}
\label{thm:multi-fold}
\thmmultifold
\end{theorem}
\newcommand{\thmprior}{
 RCTGAN~\cite{rctgan} cannot capture any $2$\textit{-fold connected} and multi-fold connected relation. SDV~\cite{sdv} and ClavaDDPM~\cite{clavaddpm} cannot capture any multi-fold connected relation. 
}
\begin{corollary}
\label{thm:multi-fold-prior}
   \thmprior
\end{corollary}

    Proof of Thm.~\ref{thm:multi-fold} and Cor.~\ref{thm:multi-fold-prior} are in App.~\ref{app:proof:multi-fold}. 
%
\noindent\textbf{CRSs in Real-World Datasets.} 
The CRSs introduced above are prevalent in real-world datasets. 
To provide a better insight, we analyzed 10 realistic RDB datasets, each containing more than 3 tables:
5 from RelBench~\cite{rdl,relbench}, and 5 from Kaggle~\cite{bike,football,social,smm,olist}\footnote{Top-ranked ones with the search keyword ``relational'' filtered by file type CSV.}.
They are of various sizes from various domains.
Table~\ref{tab:complex-data} illustrates the presence of various CRSs across these datasets. {\bf No existing model can fully capture the crucial data relations and constraints in any of the 10 datasets},
with all models being able to correctly process and satisfy the relational schema constraints for at most one dataset.
This highlights the need for a method that can effectively handle CRSs beyond the capabilities of current models.

\label{sec:schema:datasets}


\subsection{Table Order and Context Construction}
\label{sec:schema:order}
A topological order will be used during the training and generation process of IRG, so the latter tables take context from the former tables. Some topological orders would be preferred over others due to the semantic meanings of tables. For example, in the \textit{example RDB}, P preceding S makes more sense than the opposite because survey results are usually a summarized reflection of the activities a user has taken. In subsequent sections of the paper, table index $i$ refers to the index in the selected topological order, and we take the U$\rightarrow$A$\rightarrow$P$\rightarrow$S$\rightarrow$C order as the basis of discussion on the \textit{example RDB}, consistent with the indices in Fig.~\ref{fig:example-schema}. To formally define what tables are involved in the context construction of $\mathcal{T}_i$, we define an auxiliary partial-order relation as follows.
\begin{definition}
    \label{def:partial-order}

    $\mathcal{T}_i$ \textit{\textbf{affects}} ($\prec$) $\mathcal{T}_j$ if and only if: (a) $i < j$; and (b) in the sub-graph of the DAG of $\mathcal{D}$
    containing $\mathcal{T}_1, \dots, \mathcal{T}_j$, 
    $\mathcal{T}_i$ and $\mathcal{T}_j$ are connected (regardless of edge directions).
\end{definition}
\newcommand{\thmpo}{
Affects-or-equal is a well-defined partial order.
}
\begin{proposition}
    \label{thm:partial-order}
    \thmpo
\end{proposition}
\ifpreprint
The proof of Prop.~\ref{thm:partial-order} is in App.~\ref{app:proof:partial-order}.
\fi
    The affects-or-equal relation on the \textit{example RDB} is identical to the greater-or-equal of the table indices. 
    In particular, step-parent affects step-child (e.g., A $\prec$ S even in the absence of $\phi_{21}$).

\label{sec:schema:context} 

In order to capture inter-table relations more comprehensively, we must design our algorithm that breaks some constraints of $\mathcal{A}$ in Thm.~\ref{thm:multi-fold}. Most existing RDB generators have $L=2$ in Thm.~\ref{thm:multi-fold} with $f_1$ to augment tables with a context based on FKs, and $f_2$ applying an ML model to learn data distribution~\cite{sdv,clavaddpm}. Input and output for ML models are relatively fixed compared to context construction, so we can break the constraints stated in Thm.~\ref{thm:multi-fold} in $f_1$. We do so by taking information from the context constructed based on \textit{all} tables affecting the current table (recall Def.~\ref{def:partial-order}). 


Tables with parents must have FKs on them ($K_i\ge1$). FKs of a table can be ordered based on their semantic importance. The first FK is special, giving the \textit{condition} (recall that we design IRG under the \textit{conditional} generation paradigm). To construct the context using the first FK, we need to first define some auxiliary terms. 
\begin{definition}
    \textit{\textbf{Degree}}s on FK $\phi_{ik}$, denoted $d_{ik}\in\mathbb{N}^{|\mathcal{T}_{F_{ik}}|}$, are, for each row in $\mathcal{T}_{F_{ik}}$, the row counts in $\mathcal{T}_i$ sharing the same key values.
\end{definition}

Degrees are child row counts per parent row.
For instance, $d_{41}$ of the \textit{example RDB} means how many surveys each user has taken.
\begin{definition}
    \textit{\textbf{NULL-indicator}}s on a NULL-able FK $\phi_{ik}$, denoted $\tau_{ik}$, are the Boolean values indicating whether this FK in each row of $\mathcal{T}_i$ is NULL. For non-NULL-able FKs, $\tau_{ik} = \mathbf{0}$\footnote{Boolean values are represented as 0 and 1 for \texttt{False} and \texttt{True}.}.
\end{definition}

\begin{definition}
    The \textit{\textbf{aggregated}} $\mathcal{T}_i$, denoted $\widetilde{\mathcal{T}_i}$, is the result of aggregating $\mathcal{T}_i$ grouping by attributes $\mathcal{K}_{i1}\subseteq\mathcal{C}_i$ (i.e., the first FK).
\end{definition}
\ifpreprint
The aggregation function used to define an aggregated table aims to summarize information (e.g., mean and standard deviation, see details in App.~\ref{app:irg:agg}) from $\mathcal{T}_i$ attached to each row in $\mathcal{T}_{F_{i1}}$. 
\fi
\begin{definition}
    The \textit{\textbf{extended}} $\mathcal{T}_i$, denoted $\widehat{\mathcal{T}_i}$, is $\mathcal{T}_i$ with information from tables affecting it concatenated as additional attributes.
\end{definition}

\begin{figure}[t]
    \centering
    \includegraphics[width=0.9\linewidth, trim={0 4pt 0 0}, clip]{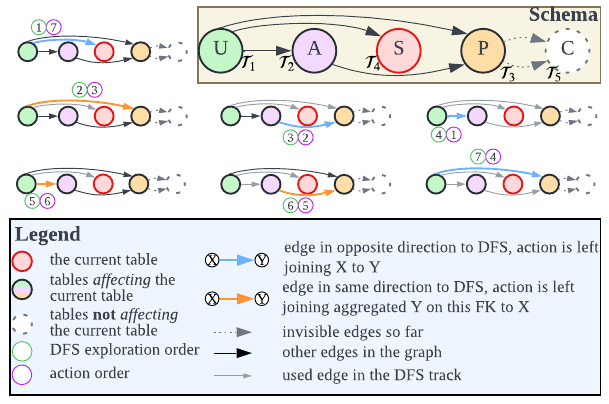}
    \vspace{-0.5em}
    \caption{Process of constructing the extended table S in the \textit{example RDB}. This algorithm is a DFS starting from table S to all tables affecting it, regardless of edge directions. Actions are triggered whenever backtracking happens.   
    Multi-fold relations are introduced in the \nth{4} and \nth{6} DFS steps.
    }
    \label{fig:join}
\end{figure}

The construction of extended tables is essentially a depth-first-search (DFS)-like traversal~\cite{dfs} on the subgraph involving all tables with smaller or equal indices (i.e., $\mathcal{T}_1,\dots,\mathcal{T}_i$). By a graph traversal starting from $\mathcal{T}_i$, the relation to all tables affecting it can be captured. This DFS approach also breaks the \textit{topological order} constraint in Thm.~\ref{thm:multi-fold}. Fig.~\ref{fig:join} illustrates the algorithm on table S in the \textit{example RDB}.
\ifpreprint
Details and complexity analysis are in App.~\ref{app:irg:join}.
\fi

\begin{definition}
\label{def:further-extended}
    The \textit{\textbf{auxiliary extended}} $\mathcal{T}_j$ up until $\mathcal{T}_i$, $j>i$, denoted $\widehat{\mathcal{T}_{ji}}$, is $\mathcal{T}_j$ extended using the same algorithm for $\widehat{\mathcal{T}_i}$ but starting from $\mathcal{T}_j$, so that additional attributes include information from all tables with indices smaller than $i$ connected to $\mathcal{T}_j$.
\end{definition}


The auxiliary extended table $\widehat{\mathcal{T}_{ji}}$ is typically a byproduct when constructing the extended table $\widehat{\mathcal{T}_i}$. In particular, auxiliary extended tables from parent tables (i.e., $\widehat{\mathcal{T}_{F_{ik}i}}$) are used widely in IRG as an augmented version of the parent table to introduce more context, such as siblings and step-parents, to the current table.

Notations introduced above are summarized in 
App.~\ref{app:notation}.

\input{algos/generate}

%% file: algos/generate.tex
\begin{algorithm}[t]
    \small
    \caption{Generating One Table ($\mathcal{T}_i$) in IRG}\label{alg:table}
    \DontPrintSemicolon
    \KwIn{
    $i$\Comment{The current table index}\;
    $\mathcal{T}_1,\mathcal{T}_2,\dots,\mathcal{T}_{i-1}$ \Comment{Previous tables, including their variants}\;
    $\Phi_1,\Phi_2,\dots,\Phi_i$ \Comment{Relevant FKs}\;
    }
    \KwOut{$\mathcal{T}_i$}
    \ForEach{$\phi_{ik}\in\Phi_i$}{
        $\widehat{\mathcal{T}_{F_{ik}i}} \gets $ \texttt{ExtendTill}$(\mathcal{T}_{F_{ik}}, \mathcal{T}_i)$\Comment{Need no info from $\mathcal{T}_i$}\;
        $d_{ik} \gets $ generated from $\widehat{\mathcal{T}_{F_{ik}i}}$\Comment{Generate degrees}
    }
    $\widetilde{\mathcal{T}_i}\gets$ generated from $\widehat{\mathcal{T}_{F_{i1}i}}$\Comment{Generate aggregated information}\;
    $\widehat{\mathcal{T}_i}\gets$ generated from $[\widehat{\mathcal{T}_{F_{i1}i}};\widetilde{\mathcal{T}_i}]$ given $d_{i1}$\Comment{Generate actual values}\;
    \ForEach{$\phi_{ik}\in\Phi_i\setminus\{\phi_{i1}\}$}{
        \If{$\phi_{ik}$ is nullable}{
            $\tau_{ik}\gets$ generated from $\widehat{\mathcal{T}_i}$\Comment{Generate NULL-indicator}
        }
        $\widehat{\mathcal{T}_i}\gets$ mapped to parent based on $d_{ik},\widehat{\mathcal{T}_{F_{ik}i}},\tau_{ik}$\;
        \Comment{Match FK attribute values}\;
        $\widehat{\mathcal{T}_i}\gets$ replaced relevant attributes using parent table's values\;
    }
    $\mathcal{T}_i\gets\pi_{\mathcal{\mathcal{C}}_i}(\widehat{\mathcal{T}_i})$ \Comment{Project to actual attributes}
\end{algorithm}

%% file: sections/4_irg.tex
In this section, we introduce how each table is generated by 5 major components to address different CRS challenges and the details of the 5 components respectively.

\begin{figure*}
    \centering
    \includegraphics[width=0.95\linewidth,trim={0.2em 0.5em 0.2em 0.2em},clip]{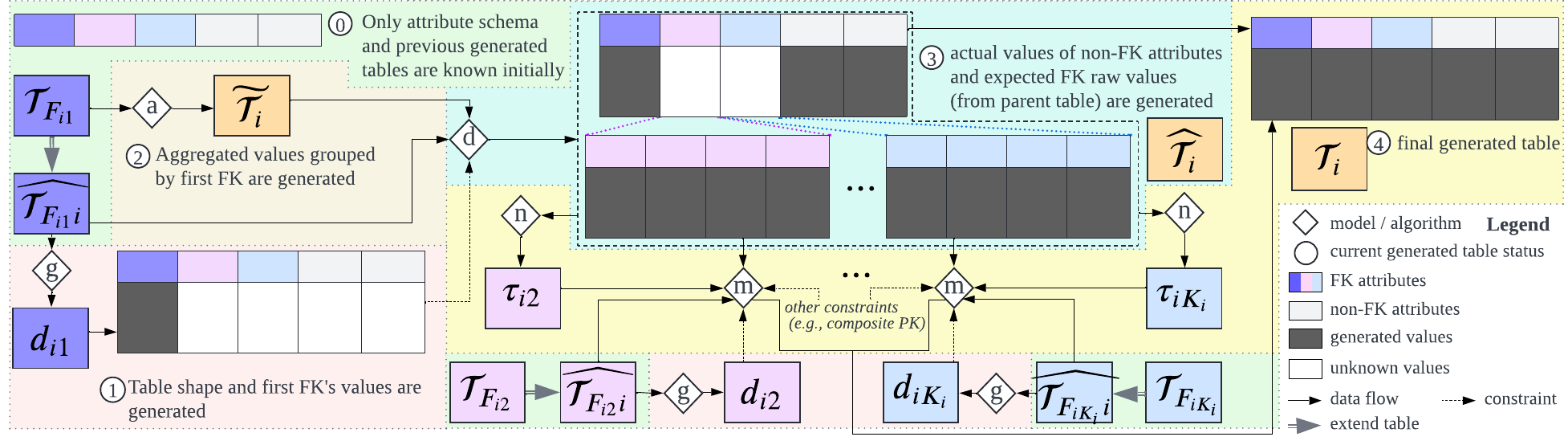}
    \vspace{-0.8em}
    \caption{Process of generating a single table (Algo.~\ref{alg:table}). 
    The 5 core components (in diamonds) are:
    degree generation (g), NULL-indicator generation (n), aggregated information generation (a), actual values generation (d), and FK attributes matching (m). 
    }
    \label{fig:irg-algo}
    \vspace{-1em}
\end{figure*}

Any table $\mathcal{T}_i$ is generated by obtaining the extended version $\widehat{\mathcal{T}_i}$ first and extracting $\mathcal{T}_i$ from $\widehat{\mathcal{T}_i}$ by simple projection.
Tables without parents have $\mathcal{T}_i=\widehat{\mathcal{T}_i}$, so they can be generated by any tabular generation model~\cite{privbayes,tablegan,ctgan,ctabgan,great,realtabformer,tabddpm,tabsyn,taegan}. 
To generate tables with parents, $\widehat{\mathcal{T}_i}$ is built up gradually, with all table variants defined in Sec.~\ref{sec:schema:order} also constructed during the process.
The overall process of generating one table is shown in Algo.~\ref{alg:table}. Fig.~\ref{fig:irg-algo} visualizes the 4 major steps of generating a single table using 5 core components: 1) ``g'' generates $d_{ik}$ determining table shapes; 2) ``a'' generates $\widetilde{\mathcal{T}_i}$; 3) ``d'' generates $\widehat{\mathcal{T}_i}$, where FKs are replaced by expected values of the parent; 4) $\tau_{ik}$ generated by ``n''
together with $d_{ik}$ and other schema constraints, are used by ``m'' 
to match each row to the parent table for FK attributes' values.
Table~\ref{tab:components} summarizes the 5 core components of IRG, whose details are elaborated in subsequent subsections.

\begin{table*}[t]
    \centering
    \caption{Overview of the five core components of IRG.
    }
    \label{tab:components}
    \vspace{-1.2em}
    \setlength{\tabcolsep}{3pt}
    \resizebox{\linewidth}{!}{
    \input{tables/components}
    }
    \vspace{-1em}
\end{table*}

Since IRG is modular, such that each component's implementation is free to use differentially private (DP) ones~\cite{dp-component1,privbayes,dp-component2,dp-component3,dpcomponent4,dp-component5}, we can have DP-IRG formalized in Theorem~\ref{thm:privacy}, constructed by DP composition rules~\cite{dp-theory,dp-composition}. The proof is in App.~\ref{app:proof:privacy}.

\newcommand{\thmprivacy}{
    \textbf{DP of IRG.}
    Given the RDB $\mathcal{D}=\{\mathcal{T}_1,\dots,\mathcal{T}_M\}$, where indices indicate the selected topological order to execute IRG.
    Suppose each component is $(\varepsilon_x,\delta_x)$-DP where $x$ indicates the the core component (recall Table~\ref{tab:components}) (e.g., the degree generation is $(\varepsilon_g,\delta_g)$-DP, except for matching, which is not DP). We add another abbreviation ``s'' indicating standalone tabular generation model. Suppose it has $M_s=|\{\mathcal{T}_i\mid K_i=0\}|$ tables without parents, and $M_t=|\{\mathcal{T}_i\mid K_i\ne0\}|$ tables with parents. Let $K$ be the total number of FKs and $K_n$ be the total number of NULL-able FKs. Then, the entire IRG process is $(\min\{M_s\varepsilon_s,\varepsilon_g+\varepsilon_a,\varepsilon_d,\varepsilon_g+\varepsilon_n\}, M_s\delta_s+M_t\delta_a+K\delta_g+K_n\delta_n)$-DP.
}
\begin{theorem}
\label{thm:privacy}
\thmprivacy
\end{theorem}
\subsection{Degree Generation}

The degrees determine the number of child rows per parent row, and hence also the \textit{shape} (sum) of the current table. 
Meanwhile, the degrees should correlate well with the parent table's content. 
The core of degree generation is regression, such that any tabular regression model can be used. However, degrees are constrained to be natural numbers within a specific range (e.g., $\ge1$ for total participation, and $\le1$ for uniqueness), and the sum determines the shape of the current table. 
We tackle these constraints by binary-searching a rounding threshold (ceil vs. floor) using values clipped to the expected range to satisfy the sum constraint.
By controlling the sum, we can also scale the generated tables flexibly.

Note that a degree of 0 typically has a special semantic meaning. 
For instance, 0-degree in $d_{21}$ means the user has never organized an activity, whose semantic difference to 1-degree is larger than the difference between degrees 1 and 2.
To preserve this distinction, we specifically control the proportion for 0-degrees.
Unlike prediction tasks, which prioritize accuracy (i.e., minimum error), the focus of generation tasks is on distribution.
To address this, we apply quantile transformation 
separately on input and output,
on top of classical regressors to better capture the distribution.
\ifpreprint
The full process is shown in Algo.~\ref{alg:deg} in App.~\ref{app:irg:degree}.
\fi

\subsection{NULL-indicator Generation}

The NULL-indicators are essential in NULL-able FKs. 
The core of NULL-indicator generation is binary classification. Similarly to degrees, the objective in this generation task is to best capture the underlying distribution. Therefore, we adjust the prediction threshold between the positive and negative classes to ensure that the class distribution is maintained.

\subsection{Aggregated Information Generation}
\label{ssec:aggregated}
Aggregated information contains a summary of each parent row's corresponding child rows of the first FK, including its other FKs and non-FK attributes, and is used to guide the actual values generation.
The generation of aggregated information from its first FK's parent table is essentially a conditional tabular generation task. The conditions are the parent rows, which are usually multivariate and continuous, instead of the class labels. Therefore, among all tabular generation models, only those specifically designed for handling parent-child relations~\cite{realtabformer,clavaddpm,rctgan}, or those that can be configured to generate feature after feature~\cite{great,taegan,tabsyn} are applicable.

\subsection{Actual Values Generation}

Actual values of the current table, where FK attributes (excluding the first FK) are replaced by expected actual values in the corresponding parent table ($\widehat{\mathcal{T}_{F_{ik}i}}$), are then generated.
This task is essentially multivariate sequential data generation with static conditions, which can be captured by conditional timeseries models~\cite{timegan,doppelganger}, but an additional length (degree) constraint should be enforced. The static conditions include information obtained from the parent table as well as generated aggregated information. We expect input from the parent table to guide the model in preserving the necessary inter-table relationships. While the aggregated information is not explicitly enforced, it is anticipated to provide sufficient guidance, thereby enhancing the quality of the generated data.

The sequential data to be generated is the actual data in the table. 
A sequential instead of a row-based model is used because rows with the same parent usually have inherent relations. For instance, if subsequent activities a user takes have a minimum interval (\textit{example RDB}), it can be captured by a sequential but not row-based model~\cite{rctgan,clavaddpm}. 
If a local sequential ID exists, its first values in each sequence are extracted in the aggregated information, and the differences between consecutive steps replace the actual values.
For example, in table S of \textit{example RDB}, the local sequential ID is \texttt{Year} (\texttt{Y}), and sequences are defined by grouping the table by \texttt{UserID} (\texttt{UID}) and sorting each group by \texttt{Y}. \texttt{Y} of the first survey is extracted as aggregated information, and consecutive survey \texttt{Y} differences replace raw \texttt{Y} values. Thus, by controlling \texttt{Y} differences to be always positive, the uniqueness of the \texttt{UID}-\texttt{Y} pair is guaranteed. 
This provides the solution for CRS where an FK overlaps with a composite PK, which has never been solved in previous works.

\subsection{FK Attributes Matching}
\label{sec:irg:match}

\begin{figure*}[t]
    \begin{minipage}{0.4\linewidth}
            \centering
            \includegraphics[width=\linewidth]{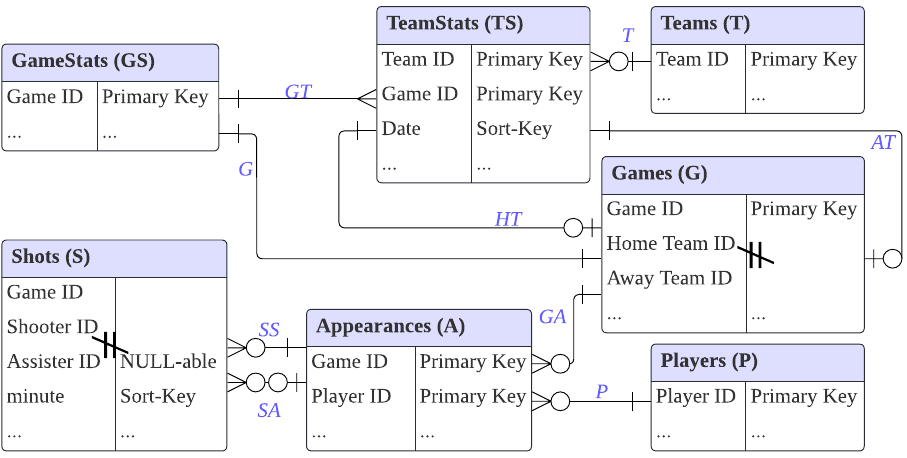}
            \vspace{-2em}
            \caption{Relational schema for the football dataset~\cite{football}. Abbr. of FKs are marked in italic blue letters, and non-self-match constraints are also marked with an inequality sign.}
            \label{fig:case-schema}
            \vspace{-1em}
    \end{minipage}
    \hfill
    \begin{minipage}{0.33\linewidth}
            \centering
            \captionof{table}{Average violation rates and violated constraint counts of CRSs. Three major schema types are checked: uniqueness of composite PKs (P), correctness of FKs (F), and correctness of non-self-matches (M). Exact 0 error is expected. Any violation is highlighted in red. 
            }
            \vspace{-1em}
            \label{tab:schema-accuracy}
            \resizebox{\linewidth}{!}{
            \input{tables/schema-accuracy-summary}
            }
    \end{minipage}
    \hfill
    \begin{minipage}{0.22\linewidth}
            \centering
            \captionof{table}{Average degree-related evaluation metrics. Columns ``K'', ``N'', and ``S'' stands for K-S statistics of distribution of degrees, difference of NULL-ratio of FKs, and the relative difference of table sizes. All values are optimal at 0. 
            }
            \label{tab:deg-summary}
            \vspace{-1em}
            \resizebox{\linewidth}{!}{
            \input{tables/deg-summary}
            }
    \end{minipage}
    \vspace{-1em}
\end{figure*}
FK matching resolves the values for FK attributes, excluding the already settled first FK (i.e., $\bigcup_{k=2}^{K_i}\mathcal{K}_{ik}\setminus\mathcal{K}_{i1}$). These FK attributes have been generated as the expected parent table values (i.e., $\widehat{\mathcal{T}_{F_{ik}i}}$) in the previous step. The actual FK values are settled by finding the corresponding parent key whose values match the expected values the best.
The core for FK matching is the finding of closest distance points, with constraints from FKs, PKs, and degrees. The simplest case is matching with only \textit{degrees} constraint. This can be modeled as 
a min-cost max-bipartite-matching (MCMBM) problem~\cite{hungarian}, where the cost is the pairwise distance, each degree corresponds to one vertex, and the bipartite graph is complete. A complete matching with minimum cost is an ideal FK matching. 
FK constraints can make some matches invalid. For example, in the \textit{example RDB}, the FKs on table C enforce that the \texttt{AID} should be the same for the two overlapping FKs, and the \texttt{UID}s for both FKs should be different (i.e., non-self-match). We refer to these collectively as \textit{valid-pair} constraints. 
MCMBM solves the case by connecting only valid pairs following the constraints. 
For cases where an additional \textit{uniqueness} constraint from composite PKs on the FKs to be matched exists, a general min-cost max-flow (MCMF) problem, instead of MCMBM, can be used.
\ifpreprint
More details of the graph modelings can be found in App.~\ref{app:irg:match}, including the graph visualizations in Fig.~\ref{fig:bipartite}-\ref{fig:mcmf}.
\fi

Although the MCMF problem is solvable, only a resulting flow of $n_i-\|\tau_{ik}\|_1$ is a valid FK matching with table shape integrity (i.e., all degrees are utilized), which is expected but does not always exist. 
\ifpreprint
In such a case, we modify the degrees generated without modifying the distribution much by reassigning a small portion of degrees to different parent rows (see App.~\ref{app:irg:match:time}). 
\fi
This solves most cases. As a last resort if degree reassignment does not enable a valid solution, we sacrifice table shape integrity to ensure schema validity.


%% file: tables/components.tex
\begin{tabular}{p{0.22\linewidth}p{0.25\linewidth}p{0.3\linewidth}p{0.22\linewidth}p{0.3\linewidth}}
    \toprule
    Name (Abbreviatio in Fig.~\ref{fig:irg-algo}) & Input & Output & Task & Complex Relational  Schema Addressed \\
    \midrule

    Degree generation (g) & $\widehat{\mathcal{T}_{F_{ik}i}}$ & $d_{ik}$, corresponding to each row in input & natural number regression with sum constraint & uniqueness \& total participation of one FK, table shape integrity, flexible table size\\\hline
    NULL-indicator generation (n) & $\widehat{\mathcal{T}_i}$ & $\tau_{ik}$, corresponding to each row in input & binary classification & FKs with NULL \\\hline
    Aggregated information generation (a) & $\widehat{\mathcal{T}_{F_{i1}i}}$ & $\widetilde{\mathcal{T}_i}$, corresponding to each row in input & multivariate conditional tabular data generation \\\hline
    Actual values generation (d) & $[\widehat{\mathcal{T}_{F_{i1}i}};\widetilde{\mathcal{T}_i}]$ as static conditions, $d_{i1}$ as lengths, and a sorting order attribute if present &  $\widehat{\mathcal{T}}_i$ 
    except for FK attributes not belonging to $\phi_{i1}$, and satisfying the length requirement
    & length-aware multivariate sequential data generation with static conditions & sequential data relation, table shapes integrity, PK by FK with local ID\\\hline
    FK attributes matching (m) & $\widehat{\mathcal{T}_i}$, $\widehat{\mathcal{T}_{F_{ik}i}}$, $d_{ik}$, $\tau_{ik}$, and other relevant PK/FKs & a map from row IDs of $\widehat{\mathcal{T}_i}$ to row IDs of $\widehat{\mathcal{T}_{F_{ik}i}}$ to fill other FK attributes of $\widehat{\mathcal{T}_i}$ & closest distance matching with constraints & PK by multiple FKs, overlapping FKs, non-self-match constraint \\
    \bottomrule
\end{tabular}

%% file: tables/schema-accuracy-summary.tex
\begin{tabular}{llccc>{\columncolor[gray]{0.9}}c}
\toprule
     &        &             IND~\cite{sdv-new} &                  RCT~\cite{rctgan} &                   CLD~\cite{clavaddpm} &             IRG \\
\midrule
\multirow{2}{*}{P}
     & Avg. &  0.000 &  \textcolor{red}{0.004} &  \textcolor{red}{0.015} &  0.000 \\
     & \# Vio. &  0/2 &  \textcolor{red}{2/2} &  \textcolor{red}{2/2} &  0/2 \\
\midrule
\multirow{2}{*}{F}
     & Avg. &        \textcolor{red}{0.663} &                \textcolor{red}{0.334} &                \textcolor{red}{0.321} &             0.000 \\
     & \# Vio. &             \textcolor{red}{7/9} &                     \textcolor{red}{4/9} &                     \textcolor{red}{4/9} &             0/9 \\
\midrule
\multirow{2}{*}{M}
     & Avg. &  0.000 &  \textcolor{red}{0.004} &  \textcolor{red}{0.064} &  0.000 \\
     & \# Vio. &  0/2 &  \textcolor{red}{2/2} &  \textcolor{red}{2/2} &  0/2\\
\bottomrule
\end{tabular}

%% file: tables/deg-summary.tex
\begin{tabular}{lccc}
    \toprule
     & K & N & S \\
    \midrule
    IND~\cite{sdv-new} & 0.627 &0.717 & \textbf{0.000} \\
    RCT~\cite{rctgan} & 0.280 & 0.481 & 0.037\\
    CLD~\cite{clavaddpm} & 0.651 & 0.442 & 0.520\\
    \midrule
    IRG & \textbf{0.036} & \textbf{0.000} & \textbf{0.000}\\
    \bottomrule
\end{tabular}

%% file: sections/5_experiments.tex
\begin{figure*}
    \begin{minipage}{0.6\linewidth}
        \centering
    \includegraphics[width=\linewidth]{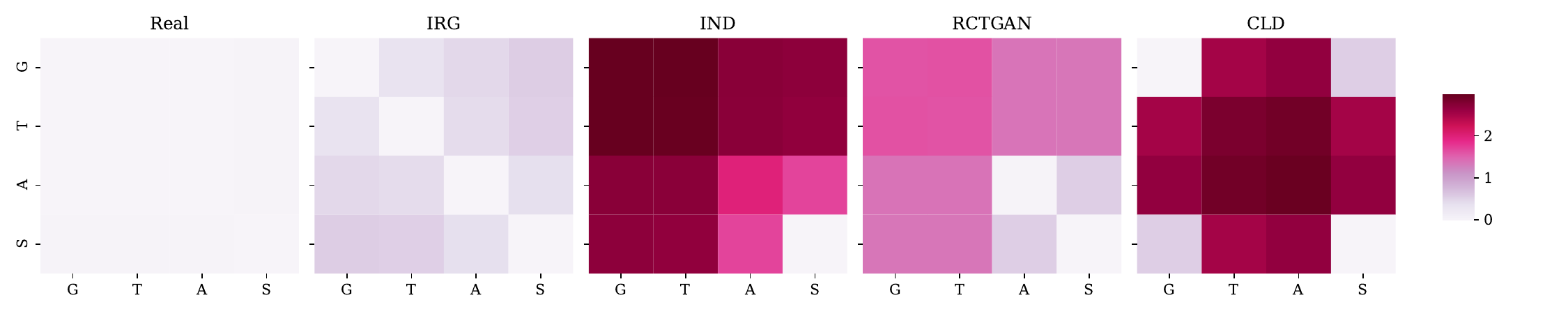}
    \vspace{-1.5em}
    \caption{The difference between goals per game computed from tables G (GS), T (TS), A, and S in $\log_{10}$ scale. Invalid values due to violated schema are replaced by $\infty$ capped to visible colors. The lighter color, the better.}
    \label{fig:goals}

        \centering
    \includegraphics[width=\linewidth]{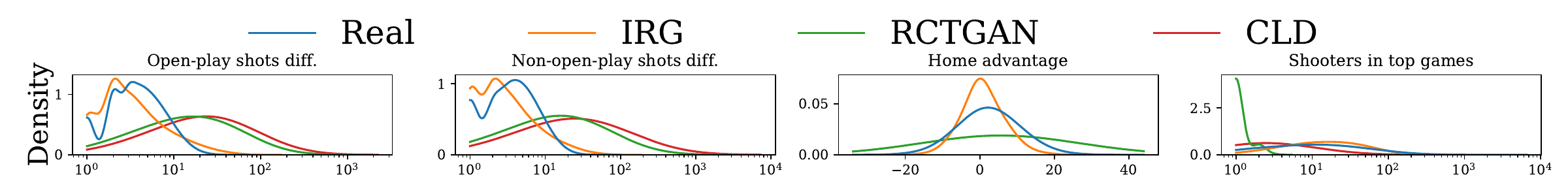}
    \vspace{-1.5em}
    \caption{Distribution of the results of SQL queries with realistic meaning in football games. Results of IND are not shown because its bad performance on FK constraints yields hardly any valid data.}
    \label{fig:queries}
    \end{minipage}
    \hfill
    \begin{minipage}{0.38\linewidth}
        \centering
    \captionof{table}{Accuracy of prediction of game results (home wins, away wins, or tie) using historical records of the teams and available public information about the expected performances, by different downstream models, trained on synthetic and tested on real. Real data's performance tested on a hold-out split is provided for reference. Best scores (highest accuracies) are highlighted in bold.}
    \label{tab:ml}
    \vspace{-0.5em}
    \resizebox{\linewidth}{!}{
    \input{tables/ml-result}

    }
    \end{minipage}
\vspace{-1em}

\end{figure*}

The advantage of IRG lies in its broad applicability across RDB datasets with CRSs. Sec.~\ref{sec:schema:datasets} already justifies the empirical advantages of IRG on real-world datasets, as baseline models are \ul{not even applicable} in most cases.
In this section, we present experimental results on a dataset modified through feature engineering to make baseline models compatible.
The experiment shows that while some tricks hack baseline models to process the data, they do not address the fundamental challenges of CRSs that IRG does. 

\subsection{Experimental Setup}
\label{sec:exp:setup}

\noindent\textbf{Baselines.}
We compare IRG with classical SDV with hierarchical modeling algorithm (HMA)~\cite{sdv}, modified SDV with independent relational modeling (IND)~\cite{sdv-new}, RCTGAN (RCT)~\cite{rctgan}, and ClavaDDPM (CLD)~\cite{clavaddpm}. 
\ifpreprint
Details can be found in App.~\ref{app:exp:baseline}.
\fi

\noindent\textbf{Dataset.}
We use the Kaggle football RDB dataset~\cite{football} (schema and corresponding abbreviations are visualized in Fig.~\ref{fig:case-schema}). This dataset contains composite PKs constructed by two FKs, two FKs from same parent table to same child table, FKs containing NULL values, and overlapping composite FKs. The dataset has a total of around 740k rows and 85 columns.
\ifpreprint
More details can be found in App.~\ref{app:exp:dataset}.
\fi

\noindent\textbf{IRG Configuration.}
We use XGBoost~\cite{xgboost} as the regressor and classifier for degree and NULL-indicator, respectively. We use modified REaLTabFormer~\cite{realtabformer} for generating aggregated information and actual values. The FK matching is done by a Google OR-Tools~\cite{ortools}. 
\ifpreprint
Details and justifications can be found in App.~\ref{app:irg}.
\fi

\ifpreprint
\noindent\textbf{Testbed.}
Experiments were conducted on an Ubuntu 22.04 server with an Intel i9-13900K CPU (24 cores, 32 threads), 125 GB of RAM, and an NVIDIA RTX 4090 GPU.
\fi

\noindent\textbf{Metrics.}
We mainly evaluate the quality of synthetic data based on its suitability for CRSs. 
\ifpreprint
Implementation details of the metrics can be found in App.~\ref{app:exp:constraints}-\ref{app:exp:ml}.
\fi
\begin{itemize}
[topsep=0pt, partopsep=0pt, itemsep=0pt, parsep=0pt,leftmargin=*]
    \item \textbf{Satisfaction of PK and FK constraints} is critical to the validity of the synthetic RDB data. \ul{Only violation rates of absolute 0 on the constraints means valid data.}
    \item \textbf{Density distribution of degrees and NULL-FKs} are important statistical properties of FK constraints and are essential to maintain table shape integrity. 
    The degrees' distributions are evaluated by K-S statistics~\cite{k-ks,s-ks}, and the NULL-FK distributions are evaluated by the difference in NULL ratio between the real and synthetic data. The difference between real and synthetic table sizes are also evaluated as part of FK-related distribution. \ul{For all these metrics, the closer to zero the better the result.}
    \item \textbf{Values computed based on realistic SQL queries} are also an indicative and important metric of RDB datasets' distribution and correlation, as queries are commonly used in analytics and data mining applications. \ul{The closer the values computed on the synthetic data to the real, the better quality of the synthetic data.}
    \item \textbf{Downstream ML tasks} where inputs and outputs are constructed from RDB, essentially a special case of SQL queries, are typical use cases of RDB data. Following the train-on-synthetic, test-on-real (TSTR) framework commonly used for evaluating synthetic tabular data~\cite{ctgan,ctabgan,tabsyn}, \ul{the better the performance of the ML task, the better the quality of synthetic data}.
\end{itemize}

\noindent\textbf{Other Experiments.}
We also do experiments on other datasets and ablation studies, with results in App.~\ref{app:result} due to space constraints.

\subsection{Applicability and Resource Consumption}
\label{sec:exp:resource}
\ifpreprint
Due to the complexity of the schemas of the dataset, \ul{no baseline directly supports its generation}. Fortunately, the dataset can be rendered in a valid input format to baselines by preprocessing (e.g., removing composite PKs, details in App.~\ref{app:exp:dataset:preprocessing}). 
\fi
Nevertheless, this na\"ive simplification of schema format provides no guarantee on the satisfaction of all schema constraints, with empirical verification in Sec.~\ref{sec:exp:constraints}.
Among all the baselines, HMA~\cite{sdv} is the most robust to complex key constraints as shown in Table~\ref{tab:comparison}, but suffers severely from scalability issues. Our trial of HMA failed due to its outrageous demand on memory. All other models run successfully. 
\ifpreprint
More details on the failure of HMA, computation time records of all models, and resource consumption analysis of IRG can be found in App.~\ref{app:result:time}.
\fi

\subsection{PK and FK Satisfaction}
\label{sec:exp:constraints}


Table~\ref{tab:schema-accuracy} shows the error rates of relational schema. The dataset we used covers many different cases in CRS, such as composite PKs, composite and overlapping FKs, FKs with non-self-match constraints, etc. The results show that na\"ive conversion of composite keys to singular keys does not help models maintain correct relational schema. IRG, on the other hand, is the only one capable of generating synthetic RDBs with valid CRSs. 
\ifpreprint
Detailed breakdowns are in App.~\ref{app:result:breakdown}.
\fi

\subsection{Degrees and NULL-FK Distribution}

%

Table~\ref{tab:deg-summary} shows the K-S statistics of different FKs from the synthetic RDBs generated by various models. Clearly, the distribution of IRG is much closer to real data, and this advantage is relatively stable across different degrees. Averaged on all 9 FKs, IRG outperforms the best-performing baseline on degree distribution by a significant margin. Due to the good performance of IRG on degrees, it is unsurprising that it also excels in table size integrity.
There is only one NULL-able FK, which is the assister from table S to A. IRG performs outstandingly better than all baselines.
\ifpreprint
Detailed breakdowns are in App.~\ref{app:result:breakdown}.
\fi

\subsection{Results of Queries}


\ifpreprint
\noindent\textbf{Consistency of goals from different tables.}
\fi
The total number of goals can be calculated from tables G (GS), T (GS), A, and S independently. The number of goals per game calculated in any way is expected to be consistent. Fig.~\ref{fig:goals} shows the differences between the goals calculated from different tables from the real and synthetic datasets. IRG has an obvious advantage in understanding inter-table relations. The result also demonstrates the necessity of keeping correct relational schemas in the synthetic data.
Observing the performance of IRG, it can be observed that the differences to G from T, A, and S increase as the distance of the tables to table G increases, demonstrating the error accumulation effect in relational models. The same trend is found in other baseline models in this graph, or the accuracy of the relational schemas.

\ifpreprint
\noindent\textbf{Distribution of SQL query results.}
\fi
There are many other meaningful SQL queries to analyze or capture characteristics of football games.
For example, the distribution of time lags between consecutive shots in a game typically shows a peak at zero for stoppages (e.g., set pieces, corners, and penalties), and most other values are condensed over small durations but occasionally extend to larger intervals in open-plays~\cite{rule}.
Home teams usually have a home advantage that they have a higher chance of winning, henceforth also more goals from home teams than away teams~\cite{home-adv1,home-adv2}.
Fig.~\ref{fig:queries} shows the distribution of 4 queries on the football dataset. Again, 
IRG shows an obvious advantage. Notably, for any arbitrary runnable SQL query on the dataset, only synthetic datasets satisfying schema constraints guarantee reasonable data, highlighting again the necessity of considering CRSs.

\subsection{Downstream Machine Learning Task}
\label{sec:exp:ml}


The football dataset can be used to predict future games' results. We run a 3-class classification task using the past performances of each team and predict the result of the current game: home team wins, away team wins, or tie.
The results are presented in Table~\ref{tab:ml}. Among 4 different ML models, from simple models to complex models, IRG shows the best averaged result, which shows the advantage of IRG in data mining, such as classification tasks.

%% file: tables/ml-result.tex
\begin{tabular}{lccc>{\columncolor[gray]{0.9}}c}
\toprule
Model     &                                Real &                           RCT~\cite{rctgan} &                  CLD~\cite{clavaddpm} &                  IRG \\
\midrule
DT &  $0.504_{\pm 0.006}$ &    $0.336_{\pm 0.059}$ &  $0.329_{\pm 0.104}$ &  $\boldsymbol{0.357_{\pm 0.005}}$ \\
RF &  $0.596_{\pm 0.001}$ &   $0.250_{\pm 0.000}$ &  $0.279_{\pm 0.022}$ &  $\boldsymbol{0.310_{\pm 0.004}}$ \\
XGB &  $0.581_{\pm 0.006}$ &   $0.250_{\pm 0.000}$ &  $\boldsymbol{0.349_{\pm 0.000}}$ &  $0.304_{\pm 0.000}$ \\
LGBM &  $0.597_{\pm 0.009}$ &  $0.250_{\pm 0.000}$ &  $0.304_{\pm 0.000}$ &  $\boldsymbol{0.320_{\pm 0.000}}$ \\
\midrule
Avg. & 0.570 & 0.272 & 0.315 & $\boldsymbol{0.323}$ \\
\bottomrule
\end{tabular}

%% file: sections/6_conclusion.tex
In this paper, we introduce IRG, a synthetic RDB data generator designed to handle datasets with complex relational schemas, which are prevalent in real-world RDB datasets. The design of the IRG framework naturally guarantees the satisfaction of relational schema constraints, including complex cases such as those involving composite keys. Moreover, IRG incorporates additional aspects of data correlations (e.g., sequential data) that are critical in RDBs but were overlooked in prior works. This enables IRG to learn data relations more effectively, going beyond just adhering to schema constraints.
Empirical results demonstrate that IRG not only satisfies complex relational schema constraints but also captures data correlations more accurately than existing models.

%% file: sections/7_appendix.tex
\section{Summary of Notations}
\label{app:notation}

\begin{table}[t]
    \centering
    \caption{Summary of notations for RDB data and schema.}
    \label{tab:notations}
    \vspace{-1em}
    \setlength{\tabcolsep}{2pt}
    \resizebox{\linewidth}{!}{
    \input{tables/notations}
    }
    \vspace{-1em}
\end{table}

\ifpreprint
\begin{table}[t]
    \centering
    \caption{Formal description using notations for some CRSs.}
    \label{tab:complex-notation}
    \vspace{-1em}
    \setlength{\tabcolsep}{3pt}
    \resizebox{\linewidth}{!}{
    \input{tables/complex-notations}
    }
    \vspace{-1em}
\end{table}
\fi

\ifpreprint
\input{tables/schema-notations}
\fi

Table~\ref{tab:notations} summarizes the notations introduced in Sec.~\ref{sec:schema:notation} for relational schemas definition. 
\ifpreprint
Table~\ref{tab:complex-notation} illustrates the formal description of different CRSs. 
\fi
Table~\ref{tab:notation-rel} summarizes the auxiliary table variants for context construction introduced in Sec.~\ref{sec:schema:context}.

\ifpreprint
\section{Supplementary Information and Analysis on the \textit{Example RDB}}

\label{app:example:config}

The \textit{example RDB} whose schema is introduced in Fig.~\ref{fig:example-schema} is an RDB of an activity-based social network platform. Its tables include:
\begin{enumerate}
    \item User (U): Users, where each user is uniquely identified by an ID.
    \item Activities (A): Activities, where each activity is uniquely identified by an ID. Each activity may have an organizer, which can be a user or NULL (meaning the activity is organized by the platform).
    \item Participation (P): Participation records of users in activities. 
    \item Yearly Survey (S): Yearly survey feedback from each user on the satisfaction of experience with the platform in the past year.
    \item Connection (C): Social connections built between users, that can only happen on an activity that both of them have participated in.
\end{enumerate}

To avoid ambiguity of the relational schema definition, we formalize the relational schema for the \textit{example RDB} as follows.
\begin{lstlisting}[language=SQL,basicstyle=\ttfamily\footnotesize]
CREATE TABLE User (
  UserID INT PRIMARY KEY,
  Gender VARCHAR(50),
  Age INT,
  RegistrationDate DATE
);

CREATE TABLE Activities (
  ActivityID INT PRIMARY KEY,
  Category VARCHAR(100),
  OpenDate DATE,
  OrganizerID INT NULL FOREIGN KEY 
    REFERENCES User(UserID)
);

CREATE TABLE Participation (
  UserID INT FOREIGN KEY 
    REFERENCES User(UserID),
  ActivityID INT FOREIGN KEY REFERENCES 
    Activities(ActivityID),
  RegistrationDate DATE,
  PRIMARY KEY (UserID, ActivityID)
);

CREATE TABLE YearlySurvey (
  UserID INT 
    FOREIGN KEY REFERENCES User(UserID),
  Year DATE,
  OverallRate FLOAT,
  ActivityCount INT,
  PRIMARY KEY (UserID, Year)
);

CREATE TABLE Connection (
  User1ID INT,
  User2ID INT,
  ActivityID INT,
  ConnectionType VARCHAR(50),
  PRIMARY KEY (User1ID, User2ID, ActivityID),
  FOREIGN KEY (User1ID, ActivityID) 
    REFERENCES Participation(UserID, ActivityID),
  FOREIGN KEY (User1ID, ActivityID) 
    REFERENCES Participation(UserID, ActivityID)
);
\end{lstlisting}

\begin{table}[h!t]
    \centering
    \caption{Content of notations for each table in the \textit{example RDB} (Fig.~\ref{fig:example-schema}). Attribute names are abbreviated by initials. For example, ``UI'' for ``User ID'', ``G'' for ``Gender''.}
    \label{tab:notation-example}
    \input{tables/eg-notations}
\end{table}
\begin{table}[h!t]
    \centering
    \caption{Content of notations for each FK in the \textit{example RDB} (Fig.~\ref{fig:example-schema}). Attribute names are abbreviated by initials. }
    \label{tab:notation-fk-example}
    \input{tables/egfk-notation}
\end{table}
\begin{table}[t]
    \centering
    \caption{Examples  of CRSs in the \textit{example RDB}.}
    \label{tab:eg-complex-only}
    \vspace{-1em}
    \setlength{\tabcolsep}{3pt}
    \resizebox{\linewidth}{!}{
    \input{tables/eg-complex-only}
    }
\end{table}
The actual meaning of the notations introduced in Sec.~\ref{sec:schema:notation} in the \textit{example RDB} is shown in Table~\ref{tab:notation-example}-\ref{tab:notation-fk-example}. Examples of CRSs are summarized in Table~\ref{tab:eg-complex-only}.
\fi

\section{Theory: Proofs}
\subsection{Proof of Thm.~\ref{thm:multi-fold} and Cor.~\ref{thm:multi-fold-prior}}
\label{app:proof:multi-fold}
\subsubsection{Prerequisites for the Proofs}
Alg. $\mathcal{A}$ as described in Thm.~\ref{thm:multi-fold} can be modeled a set of subgraph-vertex pairs: $\mathcal{V}=\{(G_t,\mathcal{T}_{i_t})|t\in\mathbb{N}\}$, where $G_t$ are connected subgraphs of the DAG of $\mathcal{D}$, and $\mathcal{T}_{i_t}$ be a vertex (table) in $G_t$. Potential global intermediate variables related to FKs can be fully represented as $\mathcal{V}$. $\mathcal{V}$ is initialized as $\emptyset$ in $\mathcal{A}$, and two processes are equivalent if the entailed $\mathcal{V}$ is equivalent. $(G_t,\mathcal{T}_{i_t})\in\mathcal{V}$ means the conditional distribution $p(\mathcal{T}_{i_t}\cap G_t|G_t\setminus\mathcal{T}_{i_t}))$ can be learned, where $\mathcal{T}_{i_t}\cap G_t$ means $\mathcal{T}_{i_t}$ with edges in $G_t$ where $\mathcal{T}_{i_t}$ is one end, and $G_t\setminus\mathcal{T}_{i_t}$ means the subgraph of $G_t$ without vertex $\mathcal{T}_{i_t}$ and edges with it as one end.
$\mathcal{V}$ is a set instead of a sequence due to the invariant of order constraint. Each call of $f_l$ is allowed to add zero or one or multiple elements in $\mathcal{V}$.

Let $\overline{\mathcal{T}}_i$ be the distribution of $\mathcal{T}_i$'s non-FK attributes, which can be learned independently of any FKs for all $i\in\{1,2,\dots,M\}$. The FK attributes' distribution cannot be learned independently of knowledge besides the current table $\mathcal{T}_i$ by their nature as relational attributes. Thus, all attributes of a table can be learned if $p(\mathcal{T}_i)=p(\overline{\mathcal{T}}_i,\phi_{i1},\phi_{i2},\dots,\phi_{iK_i})$ can be learned. Presented in graphs, this means for some $(G_t,\mathcal{T}_i)\in\mathcal{V}$, $\phi_{i1},\phi_{i2},\dots,\phi_{iK_i}$ are all present in the subgraph $G_t$.

By the constraint of $\mathcal{A}$ that can only accumulate parent-child-pair relations and Bayes Thm.~\cite{bayes}, the joint distribution $p(\mathcal{T}_i,\mathcal{T}_j)$ can be learned if and only if the following are satisfied:
\begin{itemize}
\ifpreprint
    \item \begin{enumerate}
        \item $p(\mathcal{T}_j)$ is learned; and
        \item All trails from $\mathcal{T}_j$ to $\mathcal{T}_i$ involving $\phi_{ik}$ for any $k\in\{1,2,\dots,K_i\}$ are learned.
    \end{enumerate} OR
    \item \begin{enumerate}
        \item $p(\mathcal{T}_i)$ is learned; and
        \item All trails from $\mathcal{T}_i$ to $\mathcal{T}_j$ involving $\phi_{jk}$ for any $k\in\{1,2,\dots,K_j\}$ are learned.
    \end{enumerate}
    \fi
\end{itemize}

Presented in the graphs, it means:
\begin{itemize}
    \item $\exists(G_t,\mathcal{T}_i)\in\mathcal{V}$ and all trails between $\mathcal{T}_i$ and $\mathcal{T}_j$ are in $G_t$ and $\phi_{ik}\in G_t,\forall k\in\{1,2,\dots,K_i\}$, OR
    \item $\exists(G_t,\mathcal{T}_j)\in\mathcal{V}$ and all trails between $\mathcal{T}_i$ and $\mathcal{T}_j$ are in $G_t$ and $\phi_{jk}\in G_t,\forall k\in\{1,2,\dots,K_j\}$.
\end{itemize}

A prerequisite for this condition is that any trail between $\mathcal{T}_i$ and $\mathcal{T}_j$ is found in a subgraph in $\mathcal{V}$ where the corresponding vertex is either $\mathcal{T}_i$ or $\mathcal{T}_j$, which means the relation between $\mathcal{T}_i$ and $\mathcal{T}_j$ via this trail is learned. For simplicity of reference in the subsequent parts, we call the maximum value of $n$ such that an $n$\textit{-fold connected} trail can be learned in some process as the ``maximum $n$'', denoted $n^*$.

The \textit{only FK} constraint of $f_l$ means that one call of it (on one FK) can learn the relation between the parent and child tables. Each call of $f_l$ (on $\phi$) can be considered as an arbitrary number of executions of the following process:
\ifpreprint
\begin{enumerate}
    \item Choose one end (parent or child) of the input FK $\phi$ as the core vertex. Let this end be $\mathcal{T}$.
    \item Take $(G_t,\mathcal{T}_{i_t})\in\mathcal{V}$ where $\mathcal{T}_{i_t}=\mathcal{T}'\ne\mathcal{T}$ is the other end of $\phi$, and let the corresponding subgraph be $G$, or take an empty subgraph to be $G$.
    \item Make a copy of $G$, and add the edge corresponding to $\phi$ to this $G$'s copy to be $G'$.
    \item Insert $(G',\mathcal{T})$ into $\mathcal{V}$.
\end{enumerate}
\fi

\begin{lemma}
\label{thm:multi-fold-one-limit}
    If $L=1$ in Alg. $\mathcal{A}$ as described in Thm.~\ref{alg:multi-fold}, then $n^*=1$.
\end{lemma}
\begin{proof}
    Proof by contradiction. Suppose $n^*\ge2$, then there must be a call of $f_l$ ($l$ is always 1 since $L=1$) that $(G',\mathcal{T})$ is inserted based on $(G,\cdot)$ such that $\exists$ a trail involving $\mathcal{T}$ in $G'$ that is \textit{2-fold connected} that does not exist in $G$. 
    Clearly, $G$ cannot be empty. To satisfy the above constraint, $G$ must contain a non-empty trail starting from $\mathcal{T}'$, where the direction from $\mathcal{T}'$ to an adjacent vertex $\mathcal{T}''$ on the trail is same as the direction from $\mathcal{T}'$ to $\mathcal{T}$ (direction on the original DAG). 
    Suppose this direction is forward, so $\mathcal{T}''$ and $\mathcal{T}$ are both child tables of $\mathcal{T}'$. There is a trail covering the edge $\phi'$ connecting $\mathcal{T}''$ with $\mathcal{T}$ in $G$ means this edge is introduced by some prior call of $f_l$. 
    Without loss of generality, let this call of $f_l$ be the first introduction of \textit{2-fold connected} trail with one end being the corresponding vertex for the pair to be inserted in $\mathcal{V}$. Then, after $\mathcal{A}$ has finished, $\mathcal{V}$ cannot contain $(G_t,\mathcal{T}'')$ such that a \textit{2-fold connected} trail including $\phi'$ is in $G_t$ because $L=1$.
    
    Consider a new order $\sigma_l'$ that skips $\phi'$ in the original order $\sigma_l$ and replaces it after $\phi$. This does not break the \textit{topological order} constraint, but a new \textit{2-fold connected} trail including $\phi'$ with $\mathcal{T}'$ as the paired vertex in $\mathcal{V}$ can be introduced, so that the \textit{invariant of order} constraint is violated.
    If this direction is backward, so $\mathcal{T}''$ and $\mathcal{T}$ are both parent tables of $\mathcal{T}'$, the proof is similar but in the reverse direction of the above.
    Therefore, if $L=1$, the $n^*=1$.
\end{proof}

\begin{lemma}
\label{thm:multi-fold-limit}
    The $n^*$ of Alg. $\mathcal{A}$ as described in Thm.~\ref{alg:multi-fold} is $L$.
\end{lemma}
\begin{proof}
    Note that the part in the Alg. not involving $f_l$ for any $l$ does not take in any FK, so they do not change $n^*$ of $\mathcal{A}$. 

    Proof by induction. The case of $L=1$ follows directly from Lemma~\ref{thm:multi-fold-one-limit}. Now suppose Lemma~\ref{thm:multi-fold-limit} holds for $L-1$ for some $L>1$.
    A call of $f_l$ in any iteration can:
    \ifpreprint
    \begin{enumerate}
        \item\label{case:empty} Take an empty $G$; and/or
        \item\label{case:prev} Take a $G$ that is produced in previous iterations; and/or
        \item\label{case:this} Take a $G$ that is produced by the current iteration.
    \end{enumerate}

    For case (\ref{case:empty}), $n^*$ by this round of element insertion in $\mathcal{V}$ is 1.

    For case (\ref{case:prev}), an additional edge can be added to any subgraph inserted during previous iterations, which means a potential increment of $n^*$ by 1. As only one edge can be added to a subgraph, the increment cannot be larger than 1.

    For case (\ref{case:this}), by similar arguments for the proof of Lemma~\ref{thm:multi-fold-one-limit}, the \textit{invariant of order} constraint means $n^*$ cannot be increased.

    Thus, by the inductive hypothesis on $L-1$, the $n^*$ on $\mathcal{A}$ is $L$.
    \fi
\end{proof}
\subsubsection{Proof of Thm.~\ref{thm:multi-fold}}
\ifpreprint
\begingroup
\renewcommand{\thetheorem}{\ref{thm:multi-fold}}
\begin{theorem}[Restated]
    \thmmultifold
\end{theorem}
\endgroup
\fi
\begin{proof}
    Proof of Thm.~\ref{thm:multi-fold} follows directly from Lemma~\ref{thm:multi-fold-limit}, by a $\mathcal{D}$ that has an $(L+1)$\textit{-fold connected} trail.
\end{proof}
\subsubsection{Proof of Cor.~\ref{thm:multi-fold-prior}}
\ifpreprint
\begingroup
\renewcommand{\thetheorem}{\ref{thm:multi-fold-prior}}
\begin{corollary}[Restated]
    \thmprior
\end{corollary}
\endgroup
\fi
\begin{proof}
    RCTGAN~\cite{rctgan} can be abstracted as an $\mathcal{A}$ with $L=2$ where both iterations are top-down (getting grand-parent and parent information respectively). Following from Lemma~\ref{thm:multi-fold-limit}, it cannot capture a multi-fold connected relation. However, the two iterations actually do not increase the maximum $n$, so RCTGAN is also not capable of capturing $2$\textit{-fold connected} relations. 
    
    SDV~\cite{sdv} can be abstracted as an $\mathcal{A}$ with $L=2$, where the first iteration is bottom-up for table augmentation, and the second iteration is top-down for data sampling. Thus, following from Lemma~\ref{thm:multi-fold-limit}, it cannot capture multi-fold connected relation. 
    
    ClavaDDPM~\cite{clavaddpm} generally follows SDV. It can be abstracted as an $\mathcal{A}$ with $L=2$, where the first iteration is bottom-up for latent learning and table augmentation, and the second iteration is top-down for training the classifiers. Thus, following from Lemma~\ref{thm:multi-fold-limit}, it cannot capture multi-fold connected relation. 

    RCTGAN and ClavaDDPM have an additional iteration for sampling, but they do not incur new relations as the algorithms can be written in a train-and-generate manner in 2 iterations.
\end{proof}
\ifpreprint
\subsection{Proof of Prop.~\ref{thm:partial-order}}
\label{app:proof:partial-order}
\begingroup
\renewcommand{\thetheorem}{\ref{thm:partial-order}}
\begin{proposition}[Restated]
    \thmpo
\end{proposition}
\endgroup
\begin{proof}
    By the reflexivity, anti-symmetry, and transitivity of the relation, we prove that this is a well-defined partial relation.
    \begin{itemize}
        \item \textbf{Reflexivity.} Obvious by ``-or-equal''.
        \item \textbf{Anti-symmetry.} If $\mathcal{T}_{i_1}\prec\mathcal{T}_{i_2}$ and $\mathcal{T}_{i_2}\prec\mathcal{T}_{i_1}$, then $i_2>i_1$ and $i_1>i_2$ based on the index constraint of affect relation, which is a contradiction.
        \item \textbf{Transitivity.} Suppose $\mathcal{T}_{i_1}\preceq\mathcal{T}_{i_2}$ and $\mathcal{T}_{i_2}\preceq\mathcal{T}_{i_3}$. If $\mathcal{T}_{i_1}=\mathcal{T}_{i_2}$ or $\mathcal{T}_{i_2}=\mathcal{T}_{i_3}$, then $\mathcal{T}_{i_1}\preceq\mathcal{T}_{i_3}$ is obvious. Otherwise, by the order index constraint, then $i_1<i_2<i_3$, and $\mathcal{T}_{i_1}$ and $\mathcal{T}_{i_2}$ are connected in the sub-graph up to $\mathcal{T}_{i_2}$ and $\mathcal{T}_{i_2}$ and $\mathcal{T}_{i_3}$ are connected in the sub-graph up to $\mathcal{T}_{i_3}$. Note that the sub-graph up to $\mathcal{T}_{i_2}$ and $\mathcal{T}_{i_2}$ is a sub-graph of the sub-graph up to $\mathcal{T}_{i_3}$, so that $\mathcal{T}_{i_1}$ and $\mathcal{T}_{i_2}$ are indeed also connected in the sub-graph up to $\mathcal{T}_{i_3}$. Thus, $\mathcal{T}_{i_1}\prec\mathcal{T}_{i_3}$.
    \end{itemize}
\end{proof}
\fi



\subsection{Proof of Thm.~\ref{thm:privacy}}
\label{app:proof:privacy}
\ifpreprint
\begin{theorem}[Restated]
\thmprivacy
\end{theorem}
\fi
\begin{proof}
    When the data space extend from the (processed) table to the entire RDB, originally $(\varepsilon,\delta)$-DP models are still $(\varepsilon,\delta)$-DP because DP measures the worst-case scenario.

    The first table, $\mathcal{T}_1$ has no parent, so the process up to generation of $\mathcal{T}_1$ is $(\varepsilon_s,\delta_s)$-DP. Without loss of generality, we assume all tables without parents are placed before other tables in the topological order in IRG, as this movement does not change the actual result of IRG (the affects relation is maintained).

    For table $\mathcal{T}_i$ where $i>1$, suppose the process up to the generation of $\mathcal{T}_{i-1}$ is $(\varepsilon_{i-1},\delta_{i-1})$-DP.
    If it has no parent, then the process up to generation of $\mathcal{T}_i$ is $(\varepsilon+\varepsilon_s,\delta+\delta_s)$-DP following the composition of DP~\cite{dp-foundation,dp-theory}. 
    If it has some parents, the process of generating degrees, aggregated information, and then using their outputs to generate actual values, has input generated with $(\varepsilon_{i-1},\delta_{i-1})$-DP process and the result is $(\min\{\varepsilon_{i-1},\varepsilon_g+\varepsilon_a,\varepsilon_d\},\delta_{i+1}+\delta_g+\delta_a+\delta_d)$-DP~\cite{dp-composition}. 
    If it has more than one parent, the process up to generating the first set of degrees, NULL-indicators, and FK matching is $(\min\{\varepsilon_{i-1},\varepsilon_g+\varepsilon_a,\varepsilon_d,\varepsilon_g+\varepsilon_n\},\delta_{i-1}+\delta_g+\delta_a+\delta_d+\delta_g+\delta_n)$-DP. Repeating the process for all parents, we get the process up to generating $\mathcal{T}_i$ is $(\min\{\varepsilon_{i-1},\varepsilon_g+\varepsilon_a,\varepsilon_d,\varepsilon_g+\varepsilon_n\},\delta_{i-1}+\delta_a+K_i\delta_g+|\{\tau_{ik}|\tau_{ik}\ne\mathbf{0}\}|\delta_n)$-DP.

    Eventually, by mathematical induction, the entire process is $(\min\{|\{\mathcal{T}_i\mid K_i=0\}|\varepsilon_s,\varepsilon_g+\varepsilon_a,\varepsilon_d,\varepsilon_g+\varepsilon_n\}, |\{\mathcal{T}_i\mid K_i=0\}|\delta_s+|\{\mathcal{T}_i\mid K_i\ne0\}|\delta_a+\sum_iK_i\delta_g+|\{\tau_{ik}|\tau_{ik}\ne\mathbf{0}\}|\delta_n)$-DP, namely, $(\min\{M_s\varepsilon_s,\varepsilon_g+\varepsilon_a,\varepsilon_d,\varepsilon_g+\varepsilon_n\}, M_s\delta_s+M_t\delta_a+K\delta_g+K_n\delta_n)$-DP.
\end{proof}

\ifpreprint
\section{Supplementary Notes on Realistic Datasets Analyzed}
\label{app:datasets}

The composite PKs reported in Table~\ref{tab:complex-data} are not necessarily the specified PK in the dataset, but still need to satisfy the non-NULL uniqueness constraint. Attributes involved in this non-NULL uniqueness constraint are usually the factual PK of the table, but many real-life datasets create an additional key attribute in all tables. Accordingly, composite FKs refer to FKs with the existence of such constraints on multiple attributes in the parent table. 
\fi

\ifpreprint
\section{Detailed IRG Introduction and Setup}
\label{app:irg}
IRG is essentially a modularized framework. This section describes the implementation of each module in this paper, and supplementary analysis and other possibilities of each.

\subsection{Pre-processing and Post-processing}
Before and after the IRG framework, additional pre-processing and post-processing operations can be deployed, such as the table augmentation steps of SDV~\cite{sdv} and ClavaDDPM~\cite{clavaddpm}, so that the generation of parent tables already contains direct hints for child tables. As such, all prior works in RDB generation under the conditional generation paradigm can be adapted into the IRG framework to improve their performance and utility. In fact, the table augmentation for hints for child tables inherent in augmented parent tables can be the auxiliary extended versions based on IRG to augment all tables from $\mathcal{T}_i$ to $\widehat{\mathcal{T}_{iM}}$, which also covers more information than prior works' pre-processing and post-processing.
In this paper's implementation, we focus on the core IRG steps and do not apply any of these pre-and post-processing steps apart from normalization.

\subsection{Aggregation Function for Aggregated Tables}
\label{app:irg:agg}


In this paper's implementation, we \textit{aggregate} a table using several attribute-level data-type-specific operations:
\begin{itemize}
    \item Proportion of top-3 frequent categorical values;
    \item Mean of numeric values;
    \item Median of numeric values;
    \item Standard deviation of numeric values;
    \item First row's values for trending numeric values.
\end{itemize}

Other attribute-level aggregation functions (e.g., auto-correlation, skewness) and inter-attribute aggregation functions (e.g., covariance like SDV~\cite{sdv}, neural-network-encoded values like ClavaDDPM~\cite{clavaddpm}) are also allowed, and they may be used collectively or selectively based on actual need.

In the case where $\tau_{i1}\ne\mathbf{0}$ (i.e., the attributes to be grouped on are NULL-able), NULL values are treated as a special FK value.

\subsection{More Details During \textit{Extended} Table Construction}
\label{app:irg:join}

\input{algos/extend}
In this section, we introduce more details regarding the Alg. for constructing the extended table. The process is formalized by the pseudocode shown in Alg.~\ref{alg:join}. The two \texttt{for} loops in \texttt{ExtendTill} explore edges in two directions, and in the body of each for loop, \texttt{ExtendTill} is called recursively. 

\subsubsection{Time Complexity}
\begin{theorem}
    \label{thm:complexity}
    Let $K$ be the maximum number of degrees of all vertices of the DAG derived from $\mathcal{D}$, regardless of edge directions. The time complexity of Alg.~\ref{alg:join} in terms of table joins is $O(K^{M-1})$.
\end{theorem}
\begin{proof}
    Considering table join operations only, Alg.~\ref{alg:join} is essentially $M$ modified DFS on subgraphs $\{\mathcal{T}_1\},\{\mathcal{T}_1,\mathcal{T}_2\},\dots,\{\mathcal{T}_1,\mathcal{T}_2,$ $\dots,\mathcal{T}_i\},\dots,\{\mathcal{T}_1,\mathcal{T}_2,\dots,\mathcal{T}_M\}$ (recall Sec.~\ref{sec:schema:order} for the table order). The modified DFS may visit repeated vertices and edges, but not repeated edges on the call stack for DFS backtracking.

    Let $T_i$ be the time complexity for the modified DFS on the subgraph involving $i$ vertices ($\{\mathcal{T}_1,\mathcal{T}_2,\dots,\mathcal{T}_i\}$). Then, $T_1=0=O(1)$, and for $i>1$, $T_i=K(T_{i-1}+1)=O(K^{i-1})$.

    The total time needed for Alg.~\ref{alg:join} is $\sum_{i=1}^MT_i=\sum_{i=1}^MO(K^{i-1})=O(K^{M-1})$.
\end{proof}
Although the Alg. incurs exponential complexity, it is not a serious efficiency bottleneck because the values for $K$ and $M$ are generally small even for complex RDBs, and table-join operations are much more efficient than other core components involving the training of ML models. Nevertheless, we still make IRG configurable for efficiency and scalability purposes to limit the exploration depth upward and downward, respectively. We use separate limits for the maximum depth upward and downward. Hence, IRG can be reduced minimally to na\"ive parent-and-child relation understanding only by an upward depth limit of 1 and a downward depth limit of 0. The depths are limited by passing additional upward and downward limit parameters to \texttt{ExtendTill} function and stop further exploration once the limits are hit.

\subsubsection{Aggregation Function for Joining}
We use a different aggregation function for \texttt{AggBy} in \texttt{ExtendTill} from the construction of the aggregated table because aggregation on raw data types expands the dimensions too much and is too time-consuming when applied recursively. 
For simplicity, we apply the mean and standard deviation directly on each dimension of the normalized encoded values.

\subsubsection{Dimension Reduction}
\label{app:irg:join:dim}
Note that $\widehat{\mathcal{T}_i}$ may become very large in terms of its total number of attributes. This makes it likely to go out-of-memory on common commodity machines, and excessively high-dimensional data is undesired for ML tasks due to the curse of dimensionality~\cite{cdim}. Thus, we apply dimension reduction on \textit{further-extended} table variants in \texttt{ExtendTill} in Alg.~\ref{alg:join}. 
We use principal component analysis (PCA)~\cite{pca} for dimension reduction of IRG. It is applied on \textit{further-extended} tables whenever they have the number of dimensions greater than 100, to be transformed to 100 dimensions.

For both the aggregated function and dimension reduction, instead of using na\"ive statistical methods, encoding based on neural networks (like ClavaDDPM~\cite{clavaddpm}'s latent variable encoding) may also be useful.

\subsection{Degree Generation}
\label{app:irg:degree}

Alg.~\ref{alg:deg} shows the end-to-end Alg. for generating degrees.

We use XGBoost~\cite{xgboost} with default parameters for the degree generation's regressor. The quantile transformers are implemented by \texttt{scikit-learn}. Tolerance is set to 0. The thresholds of valid degrees are obtained from the thresholds from real degree values.

Essentially any regressor (e.g., Random Forest~\cite{rf}, CatBoost~\cite{catboost}, TabNet~\cite{tabnet}) and conditional generative method (e.g., the group size generator of ClavaDDPM~\cite{clavaddpm}, or any model can be used for aggregated information generation) can be used. In addition, for better performance of all models, one can also tune the hyperparamters for the selected regressor. 
We chose XGBoost~\cite{xgboost} for its computation efficiency, ease of implementation, and outsatnding capability of capturing the relation between the target (degrees) and input variables (conditions from parent tables).

\subsection{NULL-indicator Generation}

We also use XGBoost~\cite{xgboost} with default parameters for NULL-indicator generation's classifier. However, instead of directly using the predicted class, we obtain the predicted logits and set the threshold for the separation of positive versus negative classes to maintain the generated class ratio similar to the ratio of fitting data.

Similarly to degree generation, any classifier (e.g., Random Forest~\cite{rf}, CatBoost~\cite{catboost}, TabNet~\cite{tabnet}) and conditional generative method (e.g., any model can be used for aggregated information generation) can be used, and their hyperparameters may be tuned for better performances.

\subsection{Aggregated Information Generation}



The aggregated information generation is essentially parent-to-child generation, for which we use an existing and easy-to-implement model handling this issue—REaLTabFormer~\cite{realtabformer}. We use the relational version of it, and extract the first row per parent row (a well-trained model should generate one row per parent row).
Some data standardization and normalization for a format more friendly to REaLTabFormer is applied. 
REaLTabFormer uses GPT2~\cite{gpt2} as the backbone architecture, whose utility is limited by sequence lengths, but REaLTabFormer's value representation requires a few tokens for each value, rendering sequences very long when the number of columns is reasonably large. In cases where a batch size of 1 still makes it out-of-memory, we use Longformer~\cite{led} as the backbone instead to tackle long-sequence issues.
For each model, we try with batch sizes 8, 4, 2, 1 until a run is successful without memory issues, or proceed to another backbone setting.
For models with fewer than 200 records (counted by number of parent rows), we train 500 epochs, and others train 100 epochs following the default setting. All other training settings follow the default setting of relational REaLTabFormer.

In practice, any tabular generative model that samples values column-by-column~\cite{great,taegan,tabsyn}, or any parent-to-child relational generative model~\cite{clavaddpm,spn,realtabformer} can be used for aggregated information generation.

\subsection{Actual Values Generation}

\input{algos/degree}


Although not explicitly mentioned, REaLTabFormer~\cite{realtabformer} uses a transformer to auto-regressively generate rows of the same parent row, which inherently supports sequential generation. For alignment of different components and simplicity of implementation, we use it for the actual values generation step.
Note that Longformer~\cite{led} does not improve the decoder sequence length significantly. Therefore, before running a model, we first run a preliminary size and sequence length analysis to see if the model's decoder sequence length is doomed to result in out-of-memory issues. If so, we apply window-based auto-regressive generation instead of all-at-once generation. In practice, this analysis is done by limiting the number of cells (degrees of the first FK multiplied by the number of columns) to 500. Training is based on a context size (in terms of rows) of 5 and a stride size of 10, but subject to changes based on actual dataset sizes.
The number of child rows generated per parent row is directly controllable in REaLTabFormer.
%

In practice, any multi-variate conditional sequential model~\cite{timegan,doppelganger} can be used for actual values generation. Nevertheless, we want to further pinpoint that in contrast to typical settings of multi-variate conditional sequential (e.g., time-series) generation, the actual values generation has specific requirements that existing methods may not focus on or have made oversimplistic assumptions about:
\begin{itemize}
    \item The lengths of the sequences vary, which is different from the common setting where the lengths are relatively fixed. Moreover, the lengths' distributions sometimes do not follow near-normal distributions, but rather, long-tail distributions. This is particularly problematic even if the model is capable of generating varying lengths because if na\"ive methods like padding are applied, the training data could be contaminated by the excessively large proportion of padded values.
    \item The number of features in both conditions and actual values is much larger than typical time-series or other similar sequential datasets, as they are constructed from the context of other tables and cumulatively can result in high-dimensional data, although it can be capped by controlling the maximum dimension using dimension reduction (recall App.~\ref{app:irg:join:dim}). Some methods could function well on data with lower dimensions but suffer from scalability and performance problems with high-dimensional data.
\end{itemize}

\subsection{FK Attributes Matching}
\label{app:irg:match}

\subsubsection{Graph Modeling}
\label{app:irg:match:graph}
Fig.~\ref{fig:bipartite}-\ref{fig:mcmf} shows the graph problem modeling of the FK matching as illustrated in Sec.~\ref{sec:irg:match}.
\begin{figure}
    \centering
    \includegraphics[width=\linewidth]{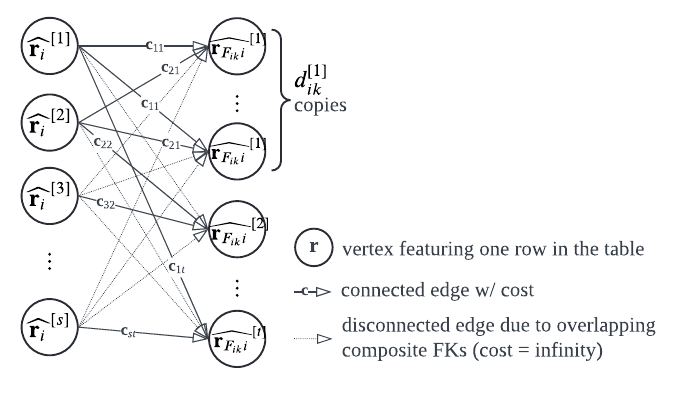}
    \caption{Min-cost max-bipartite matching (MCMBM) problem construction for FK attribute matching without \textit{uniqueness} constraint.}
    \label{fig:bipartite}
\end{figure}

Cases with constraints on the \textit{uniqueness} of certain attributes among certain matching pairs due to composite PKs constructed (fully or partially) by multiple FKs cannot be modeled as simple bipartite graphs. Nevertheless, it can be modeled as a MCMF problem by partitioning the rows of the current table $\widehat{\mathcal{T}_i}$ into some disjoint sets, as shown in Fig.~\ref{fig:mcmf}. The rows from the current table $\widehat{\mathcal{T}_i}$ can be partitioned into $q$ disjoint sets $\mathcal{S}_1,\mathcal{S}_2,\dots,\mathcal{S}_q$ such that among each set, only one element can be matched to the same row of the parent table $\widehat{\mathcal{T}_{F_{ik}i}}$. Thus, the middle layer in the constructed graph consists of pairs of the sets to parent table rows. 
Invalid pairs can be directly removed from the edges to the middle layer for the \textit{valid-pair} constraints.

\begin{figure}
    \centering
    \includegraphics[width=\linewidth]{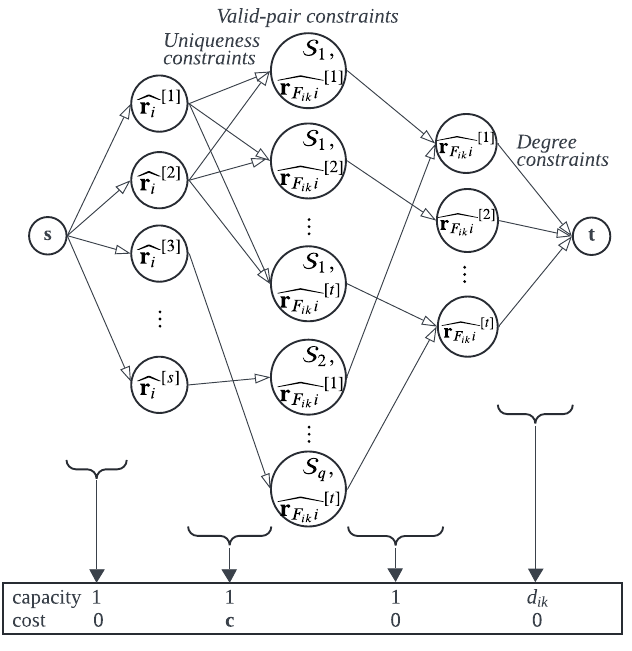}
    \vspace{-1.5em}
    \caption{Min-cost max-flow (MCMF) problem modeling for FK matching. $\mathbf{r}^{[x]}$ means the $x$-th row in table $\mathcal{T}$, where all accents and subscripts inherit from the table. $s$ and $t$ are the total number of rows in table $\widehat{\mathcal{T}_i}$ and $\widehat{\mathcal{T}_{F_{ik}i}}$ respectively. Nodes $\mathbf{s}$ and $\mathbf{t}$ represent the source and sink nodes for the max-flow problem. $\mathbf{c}$ represents the cost and is calculated by the pair-wise distance between the corresponding two rows connected.}
    \label{fig:mcmf}
\end{figure}

\begin{proposition}
    \label{thm:mcmf-validity}
    The solution of the MCMF problem is a valid solution for FK matching if an FK matching solution exists.
\end{proposition}
\begin{proof}
    For simplicity of reference, the constructed graph (Fig.~\ref{fig:mcmf}) can be considered a multi-partite, and we consider the vertices (except for source and sink vertices) as in 3 layers: first, middle, and last.

    For cases without any constraint besides the degree constraint, we can construct $\mathcal{S}_x=\{\widehat{\mathbf{r}_i}^{[x]}\}$. All sets are indeed disjoint. Each row is matched to at most one parent row by the capacity constraint between the source vertex to the first layer, and the degree constraint is fulfilled by the capacity constraint between the last layer to the sink vertex. 

    The optimal result for matching can be taken from the capacity utilization result of the MCMF solution between the first and last layer via the middle layer, where 1 means matched pairs and 0 means not matched ones.

    For cases where \textit{valid-pair} constraint exists, we can take out invalid pairs from the set of current table rows of each vertex so that the middle layer contains only valid pairs. Relevant edges between the first and second layers are removed accordingly. Thus, vertices with the same parent table row still have the sets for the current table row being disjoint, and only valid pairs can be derived.

    For cases with \textit{uniqueness} constraint, the combination of the newly matched FK with some existing attributes that are already generated or matched should be unique. Rows in the current table with relevant attributes having the same values are put in the same set. This partitioning also guarantees the resulting sets are disjoint. The \textit{uniqueness} is enforced by the capacity of 1 between the middle and the last layer.
\end{proof}

The distances are calculated by cosine distances of table rows after being encoded by the same encoder for the aggregated information generation model.

\subsubsection{Time Complexity and Efficiency Optimization}
\label{app:irg:match:time}
\begin{proposition}
    \label{thm:mcmf-complexity}
    The time complexity for both graph problems above is at least cubic w.r.t. the number of rows in the table, $n_i$.
\end{proposition}
\begin{proof}
    MCMBM problem is most efficiently solved by Hungarian Alg.~\cite{hungarian}, whose time complexity is cubic, i.e., $O(s^3)=O(n_i^3)$~\cite{hungarian-time1,hungarian-time2}.

    The general MCMF problem on a multi-partite, which is also a DAG, requires $O(F(V\log V+E))$~\cite{mcmf}, where $F=n_i$ is the total flow, $V=s+qt+t+2$ is the number of vertices, and $E=s+st+qt+t$ is the number of edges. Then, the time complexity represented by $q,s,t$ is $O(s((s+qt)\log(s+qt)+st+qt))$. We assume $O(s)=O(t)=O(q)=O(n_i)$, which is generally aligned with the cases mentioned in the proof of Prop.~\ref{thm:mcmf-validity} above. Then, the time complexity is $O(n_i^3\log n_i)$, which is also over cubic complexity.
\end{proof}

Prop.~\ref{thm:mcmf-complexity} implies a severe efficiency problem on the matching Alg. for cases where the table sizes are large. Nevertheless, we do not need a perfect solution for the minimum cost. Thus, non-perfect but efficient alternatives such as approximation Alg.s with a quadratic or even better complexity can be used instead.
%
For a large-scale bipartite matching problem, we use Sinkhorn~\cite{sinkhorn}, which has a time complexity of $O(Tn_i^2)$ for a small value $T$ with parallelism on GPU implemented by POT~\cite{pot}.
For a max flow problem, we use the cost-scaling push-relabel Alg.~\cite{mcmf-approx2,mcmf-approx} with an optimized implementation using Google OR-Tools~\cite{ortools} with a time complexity of $O(n_i^2\log n_i)$.
Some tricks to scale and convert cost values to integers are applied correspondingly to the Alg. used.
When the input for MCMBM is too large for POT that results in out-of-memory, we construct an equivalent MCMF to use Google OR-Tools~\cite{ortools} and relevant optimizations instead.

In cases where matching constraints conflict with each other, resulting in an unsolvable MCMF with the expected flow, modifying the degree constraints usually readily solves the problem. The degrees and valid-pair constraints could conflict if we stick to the generated degrees without knowledge of subsequent constraints. For example, for a specific row in the current table, the sum of degrees from all valid rows to be matched in the corresponding parent row should be greater than 1; otherwise, no valid matches could be found. If a similar pool of valid rows of the corresponding parent table is associated to many rows in the current table (usually induced by overlapping composite FKs), the sum of degrees from this pool of rows should be greater than or equal to the number of these rows in the current table. For example, to match for the \texttt{assister} in S, each row is allowed to match to a row in A with the same \texttt{game} ID. All rows in A with this \texttt{game} ID can be considered as such a valid pool, and all rows in S having the same \texttt{game} ID share the same pool. The pool may not be identical for all rows with the same \texttt{game} ID due to non-self-match constraints. The pools are detected by connected components of the valid pool for each row, where parent rows occurring in the same pool are connected. This way, most similar pools can be detected successfully, with or without additional non-self-match constraints. Then, since degrees are generated such that the sum is controlled, so the global total number of degrees are sufficient to use, but the local sum of degrees in each pool may not be enough to provide a match for all current rows sharing the same pool. Nevertheless, one degree short in one pool usually means one degree spare in another pool, where we can swap to make valid-pair constraints satisfiable without largely editing the degrees. 
The above process is summarized in Alg.~\ref{alg:deg-edit}. There are several sampling of sets in the Alg., whose success is up to the availability of content to be sampled. Nevertheless, given input $d$ with a sufficiently large sum, and by the fact that the resulting $\mathcal{G}$ is essentially a disjoint partition of $[1,2,\dots,n_{F_{ik}}]$, the sampling would always be successful, because short of one degree somewhere always mean spare of one degree somewhere else.

After degrees are edited for valid-pair constraints, the 3 types of matching constraints hardly conflict. In the extremely rare and unlikely cases where very bad parent tables are generated such that the constraints still conflict (e.g., $q=1$ while $s\gg1$ in Fig.~\ref{fig:mcmf}), the flow smaller than $n_i-\|\tau_{ik}\|_1$ is accepted, at the cost of sacrificing table size integrity.

\input{algos/deg-edit}

Even if we use the very optimized libraries for MCMF, the large dataset size easily results in out-of-memory executions. To make the Alg. executable on a normal machine, we handle rows (or disjoint groups when a uniqueness constraint is present) in batches, and update the used degrees after each batch. Each iteration (batch) is handled greedily, but each iteration typically handles a decently large amount of data, greediness in previous iterations hardly results in unsolvable next iterations.
To minimize the chance of rare cases where previous iterations' greediness affects the solvability of subsequent iterations, harder rows/groups (e.g., big disjoint groups, small valid pools) can be prioritized in earlier batches.

\subsection{General Setup for Generation One Table}
\subsubsection{Single Table Generation}
While there are many existing good-performing and efficient tabular data generation models, for simplicity of implementation, we adopt the same model for \textit{aggregated} information generation -- REaLTabFormer~\cite{realtabformer}.

\subsubsection{Example for Table Construction using Alg.~\ref{alg:table}}

\begin{figure}[t]
    \centering
    \includegraphics[width=\linewidth]{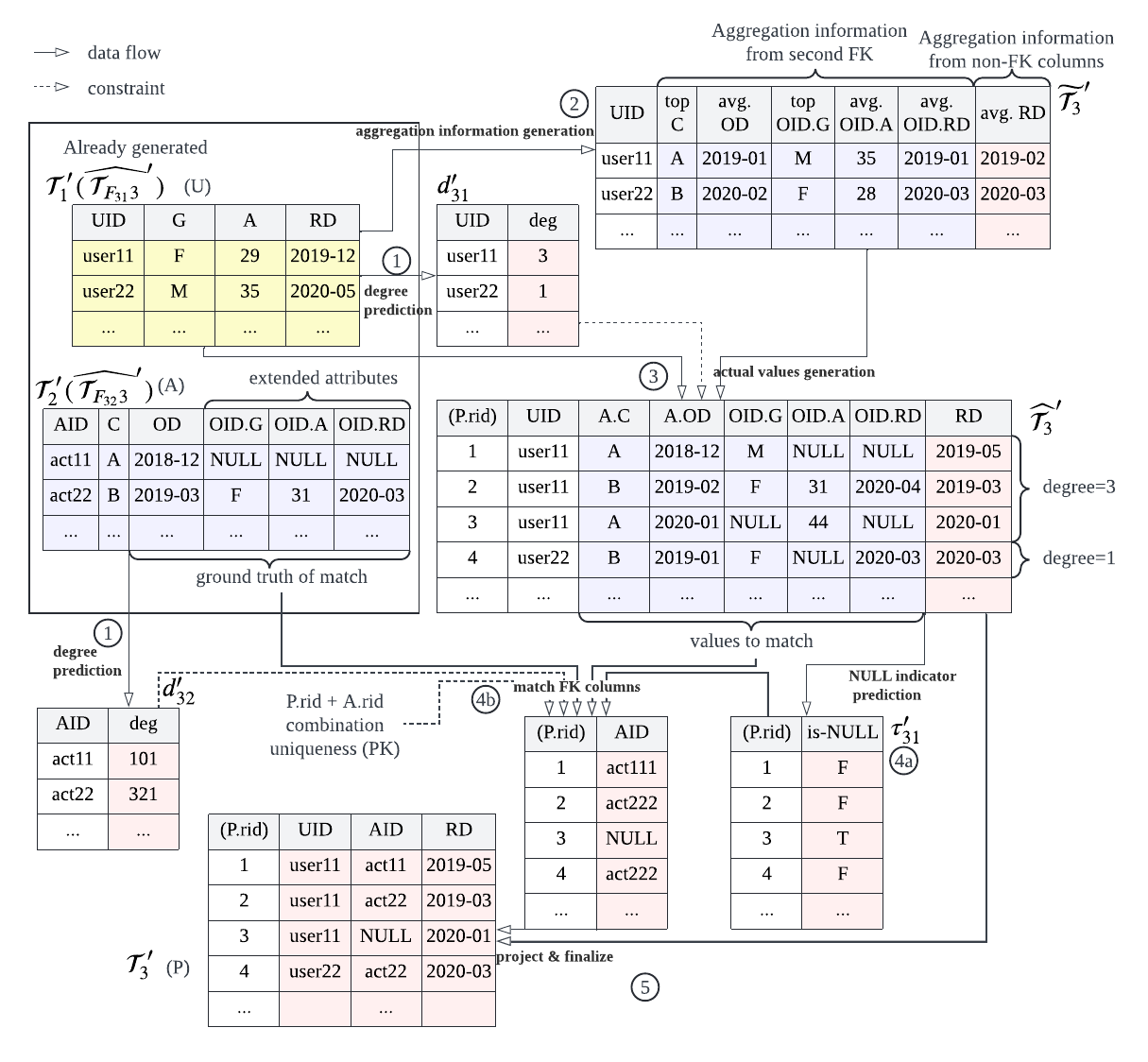}
    \caption{The actual process for constructing one synthetic table for generating table P in the \textit{example RDB}, with the variance that AID is allowed to be NULL for illustrating NULL-indicator generation.}
    \label{fig:example-process}
\end{figure}

Fig.~\ref{fig:example-process} shows the actual process of generating table P in the \textit{example RDB} following Alg.~\ref{alg:table}. To demonstrate the usage of NULL-indicator, we show the case where AID is NULL-able, which is different from the default setting.

\fi

\ifpreprint
\section{Experimental Setup}
\label{app:exp}

\subsection{Details of the dataset}
\label{app:exp:dataset}
\subsubsection{General Preprocessing}
\label{app:exp:dataset:preprocessing}
In order to make a fair comparison between all models and avoid the impact of trivial and niche engineering aspects such as the capability of reading in and handling missing values, datetime values, etc., we preprocess to remove these values to make all tables having only categorical and numeric data without missing values in all non-key attributes.
On datasets with cyclic dependencies, we manually insert intermediate tables as the feature engineering step to get rid of cyclic dependencies. Evaluations are done on the processed data. The process of removing cyclic dependencies can surely be automated, but we do not expand the method here because we focus on the generative method, and evaluations are done on the engineered versions.

\subsubsection{Football Database~\cite{football}}
\label{app:exp:dataset:football}
We use the football database~\cite{football} as the main experimental dataset. It has 7 tables: \texttt{leagues} (L), \texttt{teams} (T), \texttt{games} (G), \texttt{teamstats} (TS), \texttt{players} (P), \texttt{appearances} (A), \texttt{shots} (S). Table L has only 5 rows for 5 leagues, so we remove the table and treat it as an ordinary attribute. There is a cyclic dependency between TS and G as G introduces \texttt{game}ID while a \texttt{game}'s home and away team IDs are obtained from the TS table of team participations. To avoid cyclic dependency, we decompose G into a \texttt{gamestats} (GS) table to introduce \texttt{game} IDs and relevant statistics and a new \texttt{games} (G) table for the home and away teams. The processed schema is shown in Fig.~\ref{fig:case-schema}. The schema described in SQL is shown as follows. Non-key attributes are omitted.

\begin{lstlisting}[language=SQL,basicstyle=\ttfamily\footnotesize]
CREATE TABLE teams (
    teamID PRIMARY KEY
);

CREATE TABLE players (
    playerID PRIMARY KEY
);

CREATE TABLE gamestats (
    gameID PRIMARY KEY
);

CREATE TABLE teamstats (
    teamID,
    gameID,
    PRIMARY KEY (teamID, gameID),
    FOREIGN KEY (teamID) REFERENCES teams(teamID),
    FOREIGN KEY (gameID) REFERENCES gamestats(gameID)
);

CREATE TABLE games (
    gameID PRIMARY KEY,
    homeTeamID,
    awayTeamID,
    FOREIGN KEY (gameID, homeTeamID) 
      REFERENCES teamstats(gameID, teamID),
    FOREIGN KEY (gameID, awayTeamID) 
      REFERENCES teamstats(gameID, teamID),
    CHECK (homeTeamID <> awayTeamID)
);

CREATE TABLE appearances (
    playerID,
    gameID,
    PRIMARY KEY (playerID, gameID),
    FOREIGN KEY (playerID) REFERENCES players(playerID),
    FOREIGN KEY (gameID) REFERENCES games(gameID)
);

CREATE TABLE shots (
    gameID,
    shooterID,
    assisterID,
    FOREIGN KEY (gameID, shooterID) 
      REFERENCES appearances(gameID, playerID),
    FOREIGN KEY (gameID, assisterID) 
      REFERENCES appearances(gameID, playerID),
    CHECK (
        assisterID IS NULL OR shooterID <> assisterID
    )
);
\end{lstlisting}

\begin{table}[t]
    \centering
    \caption{Details of all FKs in the football dataset~\cite{football} and the corresponding abbreviations. Columns are represented by abbreviations from the initials of column names.}
    \label{tab:football-fks}
    \vspace{-1em}
    \resizebox{\linewidth}{!}{
    \input{tables/football-fks}
    }
\end{table}

This database contains a complex scenario of overlapping composite FKs with a non-self-match constraint and a NULL-able component. In the \text{shots} table, each \texttt{shot} made by a \texttt{player} (\texttt{shooter}) in a \texttt{game} should \texttt{appear} in the \texttt{game}, hence present in \texttt{appearances} table, and assisted by another \texttt{player} who also \texttt{appear}ed in the \texttt{game} if an \texttt{assister} exists. Moreover, the \texttt{assister} ID can be NULL, so this is also a NULL-able FK. In addition, the PK in \texttt{appearances} table, the \texttt{game}-\texttt{player} pairs, is composite and constructed by 2 FKs. This is not handled by any model other than IRG. 

To make this dataset runnable on non-IRG models, we add a new \texttt{appearance} ID column in the \texttt{appearances} table and let \texttt{shots} reference that ID to replace the composite FK. Similarly, a new \texttt{team-play} ID is added to \texttt{teamstats} and used by \texttt{games}. NULL values are replaced with auxiliary players ``NULL-x'' (multiple ``NULL''-players are used because of the dramatic difference in terms of ``the number of \texttt{shots} a \texttt{player} \texttt{assist}s'' between ``NULL'' and normal \texttt{players}). The [\texttt{game}, \texttt{shooter}]-to-\texttt{appearance} auxiliary FK will be used to recover the \texttt{game} and \texttt{shooter} ID, and [\texttt{game}, \texttt{assister}]-to-appearance auxiliary FK will be used to recover the \texttt{assister} ID only. However, this technique cannot fully resolve the overlapping composite FK problem, as there is still a high chance of FK constraint violation for these non-IRG models, as shown in the experiment results.

\subsubsection{Brazilian E-Commerce Database~\cite{olist}}
\label{app:exp:dataset:bec}
The Brazilian E-Commerce (BEC) dataset~\cite{olist} has 8 core tables, with the relational schema clearly provided on the Kaggle page of this dataset. However, investigating the actual data, we find that the zip code prefix of the geolocation table is not unique, and hence not a valid PK, likely due to data anonymization. Therefore, we process the data to use unique geolocation identifiers by aggregating the same zip code prefix to obtain the average value for geolocation information.
There is no composite FK, and no composite PK composed of 2 FKs. However, in the order items and payments tables, there are composite PKs composed of an FK with a local ID (numbers 1, 2, ...). Fortunately, they are not used by other tables as an FK, so for baselines, we simply remove these PK constraints. Nevertheless, we still evaluate them to see if baseline models are capable of capturing local sequential IDs and reporting the schema validation results, demonstrating that simplification of CRSs by removing these composite PK constraints makes the models incapable of keeping essential data patterns.

\subsubsection{Super Mario Maker Database~\cite{smm}}
\label{app:exp:dataset:smm}
The Super Mario Maker (SMM) dataset~\cite{smm} has 7 tables. The Kaggle page of the dataset provides a relational schema, but that is a loose version, and some essential data relations in the description are overlooked. Therefore, we use a tighter relational schema definition, such that: all clears, likes, and records are valid plays, and all first clear records from the course meta should be a valid clear. However, the dataset does not satisfy all these constraints, likely due to the data collection process. We clean the dataset to satisfy the constraints.
Moreover, with the new relational schema, composite PKs are introduced, for which we need to preprocess to run baseline experiments. 
Table \texttt{courses} has the only FK NULL-able, requiring additional preprocessing to decompose it into 2 tables so the NULL-able FK becomes the second FK.
In addition, this dataset has 2 FKs with NULL, which also need preprocessing for some baselines. For both scenarios, we apply the same processing steps to run baselines as the football dataset as mentioned in App.~\ref{app:exp:dataset:football}.

\label{app:exp:dataset:end}

\subsection{Implementation of Baselines}
\label{app:exp:baseline}

\subsubsection{SDV}
\label{app:exp:baseline:sdv}

We use the SDV version \texttt{1.16.2}.
SDV with IND was not provided as an end-to-end pipeline from SDV's SDK, but its corresponding training and sampling code is provided internally. We implement the end-to-end pipeline to run experiments.

\subsubsection{RCTGAN}
RCTGAN~\cite{rctgan} is generally applicable similarly to SDV, but has issues with the corner case where a table has only ID (including foreign key) columns, for which we insert a placeholder unary column to sidestep the issue.
The sampling order of tables of RCTGAN is problematic with a relatively complex schema, as it samples based on tables without parents, immediately after each of which, all its descendant tables are generated. This makes RCTGAN error-prone with tables having multiple parents. To tackle the problem, we generate all tables without parents first, without their children generated immediately after, followed by all the descendants generated together in a second loop. We manually pick a table order that does not fail the process.
An outstanding case RCTGAN fails to handle and cannot be avoided by feature engineering is (undirected) cycles in the DAG of the relational schema, due to its top-down generation strategy. This makes RCTGAN not applicable to the Super Mario Maker~\cite{smm} dataset.
The core training and generation were the same as indicated in the original source code of RCTGAN at \url{https://github.com/croesuslab/RCTGAN}.
Minor changes were applied for updated versions of certain libraries. We use default values for all parameters.

\subsubsection{ClavaDDPM}
\label{app:exp:baseline:clava}
ClavaDDPM~\cite{clavaddpm} cannot be generated even with the trick to remove composite keys mentioned above, due to the existence of FKs with the same parent and child tables, so it is not compared. Nevertheless, by inserting intermediate twin tables associated with both the parent and child for the second (and third, etc.) key of the same parent and child, the problem can be resolved. We implement the corresponding preprocessing and postprocessing steps so that experiments can be done on ClavaDDPM. The core training and generation were the same as indicated in the original source code of ClavaDDPM at \url{https://github.com/weipang142857/ClavaDDPM}.
Training configurations inherit from the provided example of \texttt{movie\_lens} dataset.

\subsubsection{Not Experimented Models}
We did not compare with relational models focusing merely on parent-child relations~\cite{realtabformer,spn}, as they are considered essentially a possible implementation of the actual values generation components or the standalone tabular data generation method instead of a complete solution for generic RDBs.

M2M~\cite{m2m}'s code link was provided in the paper, but the link is invalid.
RGCLD~\cite{rgcld} has the code available but has its preprint published only \textit{after} the first submission of this paper to KDD.
Nevertheless, not experimenting on them does not invalidate our conclusion on existing models' applicability on RDBs with CRSs.


\begin{table}[t]
\centering
\caption{Metadata and estimated column counts for football~\cite{football} tables using SDV-HMA~\cite{sdv}.}
    \resizebox{\linewidth}{!}{
    \input{tables/sdv-football}
    }
    \label{tab:sdv-football}
\end{table}

\begin{table}[t]
    \centering
    \caption{Time of training and generation on the football dataset~\cite{football} of different models in seconds.}
    \label{tab:timing}
    \input{tables/timing}
\end{table}

\begin{table*}[t]
    \centering
    \caption{Resource consumptions of core operations associated with each table on football dataset~\cite{football}. Table name abbreviations follow Fig.~\ref{fig:case-schema}.}
    \label{tab:table-resources}
    \vspace{-1em}
    \setlength{\tabcolsep}{3pt}
    \resizebox{\linewidth}{!}{
    \input{tables/table-resources}
    }
\end{table*}
\begin{table*}[t]
    \centering
    \caption{Resource consumptions of core operations associated with each FK on football dataset~\cite{football}. FK abbreviations follow Fig.~\ref{fig:case-schema} and Table~\ref{tab:football-fks}.}
    \label{tab:fk-resources}
    \vspace{-1em}
    \setlength{\tabcolsep}{3pt}
    \resizebox{\linewidth}{!}{
    \input{tables/fk-resources}
    }
\end{table*}

\subsection{Constraints Violation Calculation}
\label{app:exp:constraints}

The violation rates of the composite PK constraints are calculated by one minus the uniqueness ratio. The violation rates of FK constraints are calculated by the ratio of rows such that the non-NULL key attributes' values (or value tuples in composite cases) are found in the parent table.
The violation rates of non-self-match constraints are calculated by the ratio where the 2 FKs are equal, where inequality is expected.

\subsection{Degrees and NULL-FK Distribution}

The K-S statistics~\cite{k-ks,s-ks} is calculated using \texttt{scipy}. Although the PK and FK constraints from baseline models are already shown to result in invalid values, the degrees and NULL ratios can still be calculated following the feature-engineered constraints.

\subsection{Queries Construction and Calculation}
\label{app:exp:queries}

The numbers of goals per game from table G (GS) are calculated by the sum of home and away goals attributes. The numbers of goals from table T (TS) are computed by the sum of goals of all teams with the same game ID. The numbers of goals from table A are calculated by the sum of the sum of goals and own goals attributes grouped by game IDs. The numbers of goals from table S are calculated by the number of records with a shot result goal or own-goal grouped by game IDs. The queries in drawn in Fig.~\ref{fig:queries} are the following, but actually constructed using \texttt{pandas} operations. Rows not satisfying schema constraints are filtered out.


The duration of consecutive shots in open-play and non-open-play cases are computed using the query below.
\begin{lstlisting}[language=SQL,basicstyle=\ttfamily\footnotesize]
WITH ranked_shots AS (SELECT
  gameID, 
  minute, 
  CASE WHEN situation = 'OpenPlay' THEN 0 
    ELSE 1 END AS situation_rank,
    LAG(minute) OVER (PARTITION BY gameID ORDER BY minute, 
      CASE WHEN situation = 'OpenPlay' THEN 1 ELSE 0 END
      ) AS prev_minute,
    FROM vshots
)
SELECT 
    minute - prev_minute + 1 AS time_diff,
    CASE WHEN situation = 'OpenPlay' THEN 'open-play' 
      ELSE 'non open-play' END AS "Situation"
FROM ranked_shots
WHERE prev_minute IS NOT NULL;
\end{lstlisting}

Home advantage by shots by the difference of home and away teams is computed as follows.
\begin{lstlisting}[language=SQL,basicstyle=\ttfamily\footnotesize]
SELECT h.shots - a.shots
FROM games g
JOIN teamstats h ON g.gameID = h.gameID 
  AND g.homeTeamID = h.teamID
JOIN teamstats a ON g.gameID = a.gameID 
  AND g.awayTeamID = a.teamID
\end{lstlisting}

The goals made by top teams are calculated as follows.
\begin{lstlisting}[language=SQL,basicstyle=\ttfamily\footnotesize]
WITH game_goals AS (
SELECT 
  gameID, 
  SUM(goals) AS total_goals,
  PERCENTILE_CONT(0.75) WITHIN GROUP 
    (ORDER BY SUM(goals)) OVER () AS q3
FROM teamstats GROUP BY gameID
),
top_games AS (
  SELECT gameID FROM game_goals 
  WHERE total_goals >= q3
)
SELECT COUNT(*) FROM shots
WHERE gameID IN (SELECT gameID FROM top_games)
GROUP BY shooterID
\end{lstlisting}

\subsection{Machine Learning Task Definition}
\label{app:exp:ml}

For all games, we collect information about both the home and away teams about all their previous games and results. In particular, we include the summary of all previous games and the recent 5 games, and the attributes in table G that are known before the game starts (e.g., expected goals). Then, for each game, there are 3 possible labels: home wins, away wins, and ties. The real dataset is split into training and test sets for reference, while the synthetic dataset is always used for training and tested on \textit{all} real data. Rows with invalid schema constraints are filtered out.

We use several different ML models to showcase the utility of synthetic data under different levels of complexity of analysis—decision tree (DT)~\cite{dt}, random forest (RF)~\cite{rf}, XGBoost (XGB)~\cite{xgboost}, and LightGBM (LGBM)~\cite{lgbm}. All models are trained using default settings, and for each model, we repeat for 3 experiments and report the resulting accuracies.

\fi

\section{Supplementary Experiment Results}
\label{app:result}
\ifpreprint
\subsection{Computation Resources}
\label{app:result:time}

SDV with HMA~\cite{sdv} creates a tremendous number of auxiliary columns to capture relational information. The estimated number of columns, including these auxiliary ones in the football dataset, is shown in Table~\ref{tab:sdv-football}, obtained from the logs of SDV HMA execution. Among the 7 tables, 5 tables have more than 10,000 processed columns, and the maximum is 16 digits, which is obviously an outstanding challenge for computation machines. 
Even creating a 1-D float64 array with this size requires a memory described in PiBs, making SDV with HMA effectively inapplicable on normal computation machines, including ours.
We attempted HMA on the football dataset~\cite{football} and it failed due to memory issues after about 10 hours. Therefore, SDV-HMA results are not reported.

Table~\ref{tab:timing} shows the training (including model-specific additional preprocessing due to incapability in complex schema) and generation (including model-specific additional postprocessing to convert back to the format before preprocessing) time. No model is considered exceptionally efficient except for IND~\cite{sdv-new}, while IND fails to capture necessary inter-table constraints.
One must also note that in IRG, the computation time is very dependent on the efficiency of the actual implementations chosen at each core component. The slow generation of IRG, in particular, is due to the autoregressive generation of the backbone model, REaLTabFormer~\cite{realtabformer}, our experiments used.

We also trace the resource consumption in both time, CPU memory, and GPU memory of IRG with a more detailed breakdown. Table~\ref{tab:table-resources} and \ref{tab:fk-resources} show the resource consumption records of major table-level and FK-level operations. We also show the cumulative values with more tables processed in Table
~\ref{tab:cum-resources}. Note that the peak memory is tracked using a separate thread runs periodically, and caches from previous tasks may not be cleaned completely in actual execution, so the memory tracks can be different from the tasks launched independently. Overall, the most time-consuming and space-consuming tasks are the aggregation information and actual values' training and generation, consistent with the fact that these are the core tasks that many prior works focus on.

\begin{table*}[t]
    \centering
    \caption{The cumulative resource consumptions during IRG process on the football dataset~\cite{football}. ``T.'' and ``G.'' represents training and generation respectively. ``CPU'' and ``GPU'' means the peak memory in MBs, and ``Time'' means the time needed in seconds.}
    \begin{tabular}{ccrrrrrrr}
       \toprule
       \#tables & \#FKs & total size & T. CPU & T. GPU & T. Time & G. CPU & G. GPU & G. Time \\
       \midrule
       1	& 0 &	866B &	2566.82&	848.37&	124.38&	1250.80&	203.43	&1.00 \\
2	&0&	53.4KB&	2566.82&	854.60&	237.98&	1524.57&	203.43&	7.64\\
3&	0&	1.92MB&	3017.45&	1642.78&	1500.70&	1524.57&	1212.66&	99.08\\
4&	2	&3.56MB&	3510.61&	23484.86&	15260.90&	16390.93&	23594.33&	1974.63\\
5&	5	&3.72MB	&3510.61&	23484.86&	19545.30&	15390.93&	23594.33&	2939.78\\
6&	7&	35.67MB	&15794.42&	23484.86&	47434.56&	110179.20&	23594.33&	46792.56\\
7&	9&	68.40MB&	35012.08&	23501.77&	69698.53&	110179.20&	23594.33&	93949.62 \\
\bottomrule
    \end{tabular}
    \label{tab:cum-resources}
\end{table*}

We also show the computation consumption results of subsampled versions of the football dataset~\cite{football} in terms of the number of games to show the effect of the number of rows on the resource consumption, in Table~\ref{tab:row-ablation}.

\begin{table*}[t]
    \centering
    \caption{The resource consumption on different sub-sampled versions of the football dataset~\cite{football}. ``T.'' and ``G.'' represents training and generation respectively. ``CPU'' and ``GPU'' means the peak memory in MBs, and ``Time'' means the time needed in seconds.}
    \begin{tabular}{rrrrrrrr}
    \toprule
\#games & total size & T. CPU & T. GPU & T. Time & G. CPU & G. GPU & G. Time \\
       \midrule
         100&	0.58MB&	3325.85&	23422.20&	28240.09&	5820.60	&23457.86&	1140.51\\
1000&	5.46MB&	5119.71&	23460.33&	64360.38&	14465.64&	23443.37&	13539.89\\
12680 (full dataset)&	68.40MB&	35012.08&	23501.77&	69698.53&	110179.20&	23594.33&	93949.62\\
\bottomrule
    \end{tabular}
    \label{tab:row-ablation}
\end{table*}

On both the Brazilian E-Commence~\cite{olist} dataset and the Super Mario Maker~\cite{smm} dataset, SDV with HMA~\cite{sdv} also fails due to out of memory, similarly to the football dataset~\cite{football}. The numbers of processed columns are not as outrageously high as the football dataset, but are still as high as 7 digits, and the programs also fail after about 10 hours due to the out-of-memory issue.

\subsection{Detailed Values of Schema Satisfaction and FK-related Distributions}
\label{app:result:breakdown}
\begin{table}[t]
    \centering
    \caption{Error rates of PK constraints on composite PKs, computed by $1-\textit{the uniqueness ratio}$. For an RDB's schema to be valid, only exact 0s are accepted. Invalid values are highlighted in red. The table's name, which the composite PKs belong to follow Fig.~\ref{fig:case-schema}.}
    \label{tab:football-pk}
    \vspace{-1em}
    \input{tables/football-pk}
\end{table}
\begin{table}[t]
    \centering
    \caption{Error rates of FK constraints. Only exact 0s are accepted as valid RDB data. Invalid values are highlighted in red. FK names follow Fig.~\ref{fig:case-schema} and Table~\ref{tab:football-fks}. Some FKs in IND result in N/A values due to the bad performance in its parents, denoted as ``-''.}
    \label{tab:fk-acc}
    \vspace{-1em}
    \input{tables/football-fk-acc}
\end{table}
\begin{table}[t]
    \centering
    \caption{Error rates of non-self-match constraints. Only exact 0s mean satisfaction. Invalid values are highlighted in red. The table's name which the non-self-match constraints belong to follow Fig.~\ref{fig:case-schema}.
    There is a red 0, which means that the value is not 0 but the first 3 decimal digits are 0s.}
    \label{tab:ineq}
    \vspace{-1em}
    \input{tables/ineq}
\end{table}
\begin{table}[t]
    \centering
    \caption{K-S statistic of synthetic RDB's degrees and real data's degrees of the football dataset~\cite{football}. The smaller the value, the more similar the synthetic degrees' distribution is to the real data, featuring better synthetic data quality. FK names follow Fig.~\ref{fig:case-schema} and Table~\ref{tab:football-fks}.}
    \label{tab:deg-detail}
    \vspace{-1em}
    \input{tables/deg-dist}
\end{table}
\begin{table}[t]
    \centering
    \caption{The relative difference of the table sizes of football~\cite{football} dataset. The best (smallest) scores are highlighted in bold. Very bad values (error $\ge0.5$) are marked in red.}
    \label{tab:football-sizes}
    \vspace{-1em}
    \resizebox{\linewidth}{!}{
    \input{tables/table-sizes}
    }
\end{table}

Table~\ref{tab:football-pk}-\ref{tab:ineq} shows the satisfaction of each core CRS constraint in the football dataset~\cite{football}. The summarized result is shown in Table~\ref{tab:schema-accuracy}.
Table~\ref{tab:deg-detail} shows the K-S statistic of the degrees of all FKs. 
There is only one NULL-able FK, so Table~\ref{tab:deg-summary} shows the NULL ratio difference of this FK exactly. No further elaborated results are needed.
Table~\ref{tab:football-sizes} shows the relative difference between real and synthetic tables. IRG, due to its control of degree, generates ideal table sizes. IND also generates perfect table sizes, at the cost of very bad accuracy in relational schema constraints, because it generates tables in an independent manner. Compared to models with stronger relational consideration, IRG particularly outperforms in keeping table size integrity in tables deeper nested with foreign key constraints, while baseline models generally accumulate errors with more FK constraints. ClavaDDPM~\cite{clavaddpm} particularly performs badly in keeping table sizes on this dataset.
\fi

\subsection{Results on Brazilian E-Commerce~\cite{olist}}
\label{app:result:bec}

\ifpreprint
\begin{table}[t]
    \centering
    \caption{Error rates of composite PK constraints, computed by $1-\textit{the uniqueness ratio}$. For an RDB's schema to be valid, only exact 0s are accepted. Invalid values are highlighted in red. ``OI'' and ``OP'' refer to \texttt{order\_item} and \texttt{order\_payments} tables respectively. There is a red 0, which means that the value is not 0 but the first 3 decimal digits are 0s.}
    \label{tab:olist-pk}

\input{tables/olist-pk}
\end{table}
\fi
We mainly focus on the schema constraints satisfaction for other datasets.
There is no composite FK in the dataset, so all baselines manage to achieve 0 error in FKs, except for IND, which has an error rate of 1 for all FKs due to its fundamentally non-relational design.
Table~\ref{tab:olist-pk} shows the satisfaction of composite PKs, which are composed of an FK and local ID in this dataset. Again, IRG manages to maintain the CRS constraints while other models cannot guarantee this.
There is no non-self-match constraint on this dataset.

\subsection{Results on Super Mario Maker~\cite{smm}}
\label{app:result:smm}

\ifpreprint
\begin{table}[t]
    \centering
    \caption{Summary of error rates on FK constraints on SMM dataset~\cite{smm}. Exactly 0 error rates are expected. Violations are highlighted in red.}
    \label{tab:smm-fk-acc}
    \input{tables/smm-fk-acc}
\end{table}
\begin{table}[t]
    \centering
    \caption{Error rates of composite PK constraints on SMM~\cite{smm} datasets. Exactly 0 error rates are expected. Violations (including non-zero values in digits after the 3 digits displayed) are highlighted in red.}
    \label{tab:smm-pks}
    \input{tables/smm-pks}
\end{table}
\fi

\ifpreprint Recall App.~\ref{app:exp:baseline} that \fi HMA~\cite{sdv} and RCTGAN~\cite{rctgan} fail to run on this dataset. IND~\cite{sdv-new}, like always, has error rates of nearly 1 on non-leading FKs constraints (i.e., $\phi_{ik}$ where $k\ge2$). ClavaDDPM~\cite{clavaddpm} fails with an error rate of nearly 1 on FKs with overlapping attributes with previous FKs (first clear of \texttt{course\_meta}). In comparison, IRG satisfies all FK constraints. A summary is in Table~\ref{tab:smm-fk-acc}.
\ifpreprint
Although ClavaDDPM~\cite{clavaddpm} satisfies FK constraints better than IND~\cite{sdv-new}, the incapability in composite PK constraints makes it fail to satisfy all composite PK constraints. IND~\cite{sdv-new} is able to satisfy composite PKs by two singular FKs, but is unable to satisfy composite PKs by one composite FK. 
Detailed \fi PK satisfaction is in Table~\ref{tab:smm-pks}. Similar to other datasets, IRG showcases its outstanding ability in maintaining the CRS constraints, and other baseline models are unable to. Moreover, the fact that even with feature engineering, RCTGAN~\cite{rctgan} fails to run shows that explicitly considering CRSs is increasingly an essential demand with the increasing complexity of relational schemas.

\ifpreprint
\section{Ablation Study}
\label{app:abl}

\subsection{Each Component Plays a Role in Maintaining CRSs}
As shown in Table~\ref{tab:components}, each of the 5 components is in charge of a subset of CRSs (except for ``a’’). They together form the core design of the IRG. Although different datasets may trigger only some instead of all of the components, removal of each component would result in the incapability of model in handling corresponding CRSs. We cannot run ablation experiments, but the correspondence of the components with different CRSs shown in Table~\ref{tab:components} provides some sense of theoretical ablation on the capability of IRG with different components in handling datasets with different CRSs.

\subsection{Controlling the Distribution of Degrees and NULL Indicators}
\fi
\label{app:abl:dist}
\begin{figure*}
\begin{minipage}[b]{0.65\linewidth}
    \centering
    \includegraphics[width=\linewidth]{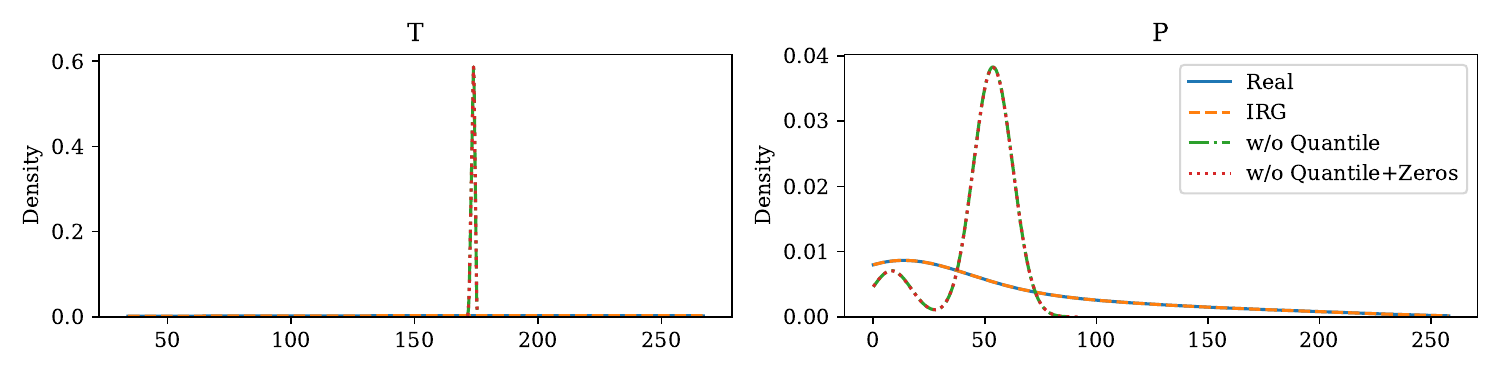}
    \vspace{-1.5em}
    \caption{Ablation by distribution of degrees of real and synthetic data.}
    \label{fig:abl-deg-q}
\end{minipage}
\hfill
\begin{minipage}[b]{0.32\linewidth}
    \centering
    \includegraphics[width=\linewidth,trim={0 0 0 2em}, clip]{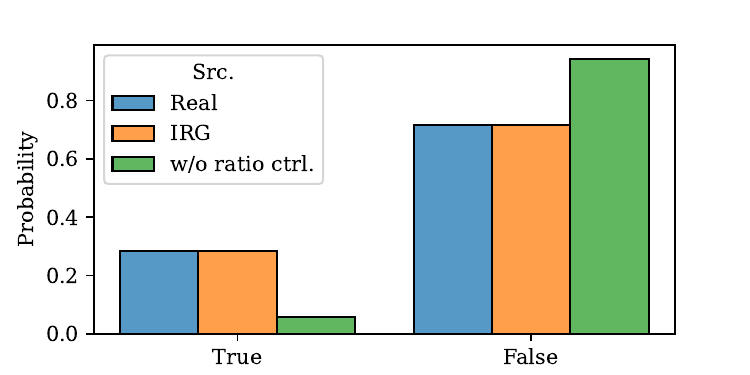}
    \vspace{-2.5em}
    \caption{Ablation on ratio of NULL FKs w/ vs. w/o ratio control.}
    \label{fig:abl-na-ratio}
\end{minipage}
\vspace{-1.5em}
\end{figure*}

\ifpreprint
\begin{table}[t]
    \centering
    \caption{Ablation study on the sum of degrees of real and generated data. ``Exp. sum'' means the expected sum of degrees (by real data), ``IRG sum'' means the generated sum of degrees by IRG, and ``w/o sum ctrl.'' means the ablation where the degree generation has no control over sum.}
    \label{tab:abl-sum-deg}
    \input{tables/abl-sum-deg}
\end{table}
\fi
The generation of degrees and NULL-indicators is based on simple regression and classification tasks, but we designed a mechanism to purposely control the distributions. This is important in a generative task. Fig.~\ref{fig:abl-deg-q} shows the comparison of the distribution of degrees generated using the original IRG design, with the quantile component removed, and further with the zero ratio control removed, respectively, on the total number of games a team participates in (T), and the total number of appearances a player has made (P). Degrees of other FKs are not visualized because the difference with or without distribution control is not obvious. Clearly, controlling the distribution explicitly results in a much better generated distribution.
Controlling the sum of degrees is also essential in maintaining the table size integrity. Even with the distribution control, if we directly output the rounded and clipped regression result as degrees, the sum of degrees could still be very different from the expected sum of degrees due to prediction errors. Table~\ref{tab:abl-sum-deg} shows the direct output rounded and clipped of degrees and the expected degrees of all 9 FKs in the football dataset~\cite{football}. It shows the necessity of controlling the sum.
For NULL indicators, similarly, the ratio of NULL and NOT NULL directly computed from the classifier may be very different from the expected ratio, while a synthetic RDB with good quality is expected to maintain a similar ratio. Fig.~\ref{fig:abl-na-ratio} shows the comparison between the proportion of the two classes with and without ratio control, re-emphasizing the importance of controlling the distribution in these seemingly simple tasks for generation.

\ifpreprint
\subsection{Treating Actual Values as Sequential Data}
\label{app:abl:seq}
\fi
\begin{table}[t]
    \centering
    \caption{Ablation study using downstream ML. Real and IRG results follow from Table~\ref{tab:ml}, ``w/o seq.'' means using a degraded tabular model for actual data generation, and ``small comp.'' means when actual data generation components are replaced by a smaller model.}
    \label{tab:abl-components}
    \vspace{-1em}
    \input{tables/ablation-component}
\end{table}
Treating actual values as sequential data instead of atomic rows allows the model to maintain sequential relations, which could be important. We replaced the sequential component of TS with an atomic row-based generation, still implemented by REaLTabFormer~\cite{realtabformer}, which could be regarded as a degraded sequential generative model, and re-evaluated the ML downstream tasks. The results before and after the change are shown in Table~\ref{tab:abl-components} (``w/o seq.''). This validates the necessity of treating rows with the same FK as sequential entities rather than atomic singular components.

\ifpreprint
\subsection{Effect of the Performance of Each Component}
\label{app:abl:comp}
\fi
As mentioned in the main paper, IRG is essentially a framework, where its performance is largely dependent on the model used in each component. Intuitively, better models used in each component, better synthetic data can be produced. We still use the ML task to do the ablation study, where we reduce the sizes of the backbone of REaLTabFormer~\cite{realtabformer} significantly by reducing the dimensions and layers. Using the original and reduced models does not make IRG incomplete, but, as shown in Table~\ref{tab:abl-components} (small comp.), a better model (we treat the larger as better here) yields a better result.
This modular characteristic of IRG is intentional, for its future extensibility using the same framework when better tabular, conditional tabular, and conditional sequential generative models are available.

%% file: tables/notations.tex
\begin{tabular}{llp{0.6\linewidth}}
    \toprule
    Group & Notation & Description \\
    \midrule
    \multirow{2}{*}{General} & $\alpha'$ & synthetic counterpart of $\alpha$ \\
    & $|\cdot|$ & cardinality, or size of a set\\
    \hline
    \multirow{9}{*}{Core} & $\mathcal{D}$ & RDB data with schema \\
    & $\mathcal{T}_i$ & $i$-th table in $\mathcal{D}$ \\
    & $\mathcal{C}_i$ & set of attributes for $\mathcal{T}_i$ \\
    & $\mathcal{P}_i\subseteq\mathcal{C}_i$ & PK attributes in $\mathcal{T}_i$ \\
    & $\Phi_i$ & set of FKs of $\mathcal{T}_i$ \\
    & $\phi_{ik}\in\Phi_i$ & $k$-th FK on $\mathcal{T}_i$ \\
    & $F_{ik}\in\{1,2,\dots,M\}\setminus\{i\}$ & parent table index for $k$-th FK on $\mathcal{T}_i$ \\
    & $\mathcal{K}_{ik}\subseteq\mathcal{C}_i$ & attributes in $\phi_{ik}$ from $\mathcal{T}_i$\\
    & $\mathcal{F}_{ik}\subseteq\mathcal{C}_{F_{ik}}$ & attributes in $\phi_{ik}$ from $\mathcal{T}_{F_{ik}}$\\
    \hline
    \multirow{4}{*}{Sizes} & $M=|\mathcal{D}|\in\mathbb{N}^+$ & number of tables in $\mathcal{D}$ \\
    & $n_i=|\mathcal{T}_i|\in\mathbb{N}^+$ & number of rows in $\mathcal{T}_i$ \\
    & $|\mathcal{C}_i|\in\mathbb{N}^+$ & number of attributes in $\mathcal{T}_i$\\
    & $K_i=|\Phi_i|\in\mathbb{N}$ & number of FKs in $\mathcal{T}_i$\\
    \hline
    \multirow{2}{*}{Index} & $i,j\in\{1,2,\dots,M\}$ & table index in $\mathcal{D}$ \\
    & $k\in\{1,2,\dots,K_i\}$ & FK index in table $\mathcal{T}_i$ \\
    \bottomrule
\end{tabular}

%% file: tables/complex-notations.tex
\begin{tabular}{p{0.65\linewidth}p{0.65\linewidth}}
    \toprule
    Case & Formal Description \\
    \midrule
    $\ge2$ FKs with same parent \& child tables & $k_1\ne k_2\in\{1,2,\dots,K_i\}$ s.t. $F_{ik_1}=F_{ik_2}$\\
    PK with $\ge2$ attributes & $|\mathcal{P}_i|\ge2$\\
    $\ge2$ FKs form a PK & $\bigcup_{k\in A}\mathcal{F}_{ik}=\mathcal{P}_i$ where $A\subseteq\{1,2,\dots,K_i\},|A|\ge2$\\
    $\ge1$ FK \& a local ID as PK & $\bigcup_{k\in A}\mathcal{F}_{ik}\subseteq\mathcal{P}_i$ where $A\subseteq\{1,2,\dots,K_i\},|A|\ge1$\\
    FK with $\ge2$ attributes & $|\mathcal{K}_{ik}|\ge2$\\
    $\ge2$ composite FKs with overlapping attributes & $k_1\ne k_2\in\{1,2,\dots,K_i\}$ s.t. $\mathcal{K}_{ik_1}\cap\mathcal{K}_{ik_2}\ne\emptyset$\\
    \bottomrule
\end{tabular}

%% file: tables/schema-notations.tex
\begin{table}[t]
    \centering
    \caption{Notations for relational schema understanding.}
    \label{tab:notation-rel}
    \vspace{-1em}
    \resizebox{0.8\linewidth}{!}{
    \begin{tabular}{cl|cl}
        \toprule
        $d_{ik}$ & degrees & $\tau_{ik}$ & NULL-indicators \\
        $\widehat{\mathcal{T}_i}$ & extended & $\widetilde{\mathcal{T}_i}$ & aggregated \\
        $\widehat{\mathcal{T}_{ij}}$ & auxiliary-extended & $\widetilde{\mathcal{T}_i}^*$ & aggregation-transformed \\
        \bottomrule
    \end{tabular}
    }
\end{table}

%% file: tables/eg-notations.tex
\begin{tabular}{l|lllll}
    \toprule
     & U ($\mathcal{T}_1$) & A ($\mathcal{T}_2$) & P ($\mathcal{T}_3$) & S ($\mathcal{T}_4$) & C ($\mathcal{T}_5$) \\
    \midrule
    $i$ & 1 & 2 & 3 & 4 & 5\\
    \multirow{2}{*}{$\mathcal{C}_i$} & UI, G, & AI, C, & UI, AI, & UI, Y, & U1I, U2I, \\
    & A, RD & OD, OI & RD & OR, AC & AI, CT \\
    $\mathcal{P}_i$ & UI & AI & UI, AI & UI, Y & U1I, U2I, AI\\
    $K_i$ & 0 & 1 & 2 & 1 & 2 \\
    $\Phi_i$ & & $\phi_{21}$ & $\phi_{31},\phi_{32}$ & $\phi_{41}$ & $\phi_{51},\phi_{52}$\\
    \bottomrule
\end{tabular}

%% file: tables/egfk-notation.tex
\begin{tabular}{l|llllll}
    \toprule
     & $\phi_{21}$ & $\phi_{31}$ & $\phi_{32}$ & $\phi_{41}$ & $\phi_{51}$ & $\phi_{52}$\\
    \midrule
    $i$ & 2 & 3 & 3 & 4 & 5 & 5\\
    $k$ & 1 & 1 & 2 & 1 & 1 & 2\\
    $F_{ik}$ & 1 & 1 & 2 & 1 & 3 & 3\\
    $\mathcal{K}_{ik}$ & OI & UI & AI & UI & U1I, AI & U2I, AI \\
    $\mathcal{F}_{ik}$ & UI & UI & AI & UI & UI, AI & UI, AI \\
    NULL-able? & Y & N & N & N & N & N \\
    \bottomrule
\end{tabular}

%% file: tables/eg-complex-only.tex
\begin{tabular}{p{0.7\linewidth}p{0.25\linewidth}}
    \toprule
    Case &  Example \\
    \midrule
    FK with NULL & $\phi_{21}$ \\
    $\ge2$ FKs with same parent \& child tables & $\phi_{51},\phi_{52}$\\
    $\ge2$ FKs with non-self-match constraint & $\phi_{51},\phi_{52}$ \\
    PK with $\ge2$ attributes & $\mathcal{P}_3, \mathcal{P}_4, \mathcal{P}_5$\\
    FK with $\ge2$ attributes & $\phi_{51},\phi_{52}$\\
    $\ge2$ FKs form a PK & $\mathcal{P}_3,\mathcal{P}_5$\\
    $\ge1$ FK \& a local ID as PK & $\mathcal{P}_4$\\
    $\ge2$ composite FKs with overlapping attributes & $\phi_{51},\phi_{52}$ \\
    Sequential relation & $\mathcal{T}_4$ \\
    Step-parent \& step-child & $\mathcal{T}_2,\mathcal{T}_4$ witho $\phi_{21}$\\
    \bottomrule
\end{tabular}

%% file: algos/extend.tex
\begin{algorithm}[t]
\small
\caption{Constructing \textit{Extended} Tables}\label{alg:join}
\DontPrintSemicolon
\KwIn{$\mathcal{D}$ with $\mathcal{T}_1,\mathcal{T}_2,\dots,\mathcal{T}_M$ and corresponding $\Phi_1,\Phi_2,\dots,\Phi_n$\;\Comment{The database and schema}}
\KwOut{$\widehat{\mathcal{T}_1},\widehat{\mathcal{T}_2},\dots,\widehat{\mathcal{T}_M}$}
\SetKwFunction{FDesc}{ExtendTill} 
\SetKwProg{Fn}{Function}{:\Comment{\normalfont Returns $\widehat{\mathcal{T}_{ji}}$}}{}
\Fn{\FDesc{$\mathcal{T}_j$, $\mathcal{T}_i$, $\Phi$}}{ 
    \Comment{$\Phi$ is the processed FK queue to avoid endless loop}\;
    $\widetilde{\mathcal{T}_j}^*\gets\widetilde{\mathcal{T}_j}$ transformed and mapped to the same size as $\mathcal{T}_j$\;
    $\mathcal{T}\gets[\widetilde{\mathcal{T}_j}^*; \widehat{\mathcal{T}_j]}$ \Comment{$j<i$ is already generated, include agg.}\;
    \ForEach{$\phi_{jk}\in\Phi_j$}{
        \If{$\phi_{jk}\notin\Phi$}{
            $\widehat{\mathcal{T}_{F_{jk}i}}\gets$ \texttt{ExtendTill}($\mathcal{T}_{F_{jk}}$, $\mathcal{T}_i$, $\Phi\cup\{\phi_{jk}\}$)\;
            $\mathcal{T}\gets\mathcal{T}\Join_{\phi_{jk}}\widehat{\mathcal{T}_{F_{jk}i}}$\Comment{Left join}
        }
    }
    children $\gets\{\phi_{i'k}|F_{i'k}=j,i'<i\}\setminus\Phi$\;
   \ForEach{$\phi_{i'k}\in$ {\normalfont children}}{
        $\widehat{\mathcal{T}_{i'i}} \gets $ \texttt{ExtendTill}($\mathcal{T}_{i'}$, $\mathcal{T}_i$, $\Phi\cup\{\phi_{i'k}\}$)\;
        $s \gets \widehat{\mathcal{T}_{i'i}}$.\texttt{aggBy}($\mathcal{K}_{i'k}$)  \Comment{Aggregation to be elaborated}\;
        $\mathcal{T} \gets \mathcal{T}\Join_{\phi_{i'k}} s$ \Comment{Left join}
   }
    \KwRet $\mathcal{T}$\;
}
\SetKwFunction{FJoin}{ExtendTable}
\SetKwProg{Fn}{Function}{: \Comment{\normalfont Returns $\widehat{\mathcal{T}_i}$}}{}
\Fn{\FJoin{$\mathcal{T}_i$}}{ 
        $r \gets \mathcal{T}_i$ \Comment{Initialized as itself}\;
        \ForEach{$\phi_{ik}\in\Phi_i$}{
            $\widehat{\mathcal{T}_{F_{ik}i}}\gets$ \texttt{ExtendTill}($\mathcal{T}_{F_{ik}}$, $\mathcal{T}_i$, $\emptyset$) \Comment{Start with empty queue}\;
            $r\gets r\Join_{\phi_{ik}} \widehat{\mathcal{T}_{F_{ik}i}}$ \Comment{Left join}\;
        }
        \KwRet $r$\;
}
\ForEach{$\mathcal{T}_i\in\mathcal{D}$}{
    $\widehat{\mathcal{T}_i} \gets $ \texttt{ExtendTable}($\mathcal{T}_i$)\;
}
\end{algorithm}

%% file: algos/degree.tex
\begin{algorithm}[t]
    \small
    \caption{Generating Degrees}\label{alg:deg}
    \DontPrintSemicolon
    \KwIn{\;
    $\widehat{\mathcal{T}_{F_{ik}i}},d_{ik},\widehat{\mathcal{T}_{F_{ik}i}}'$\Comment{Training $\mathbf{X, y}$ and testing $\mathbf{X}$ for the regressor}\;
    $S$\Comment{Expected sum of degrees after scaling}\;
    $\varepsilon$\Comment{Tolerance ratio of sum to $S$, must be 0 if $i>1$}\;
    $d_{\min},d_{\max}$\Comment{Threshold of valid degrees based on constraints}\;
    \KwOut{$d_{ik}'$\Comment{Generated degrees}}
    $\mathcal{M}\gets$ fit regressor $\mathcal{M}$ with $\mathbf{X}=\widehat{\mathcal{T}_{F_{ik}i}},\mathbf{y}=d_{ik}$\;
    $d\gets\mathcal{M}(\widehat{\mathcal{T}_{F_{ik}i}})$\Comment{Get predictions}\;
    $e\gets$ 95\% percentile from relative error $\frac{|d-d_{ik}|}{\max d_{ik}}$\;
    \Comment{Inverse confidence on $\mathcal{M}$}\;
    $w\gets\min\{2e,1\}$\Comment{Weight on $\mathcal{Q}$}\;
    $r\gets$ zero ratio in $d_{ik}$\;
    $\mathcal{Q}\gets$ fit quantile transformer with non-zero values in $d_{ik}$\;
    $s\gets\sum d_{ik}$\Comment{Get real sum of degrees}\;
    \Comment{End of training stage}\;
    $d_{ik}'\gets\mathcal{M}(\widehat{\mathcal{T}_{F_{ik}i}})$\Comment{Make model prediction}\;
    $\mathcal{Q}'\gets$ fit quantile transformer with $d_{ik}'$ \;
    $q\gets\mathcal{Q}'(d_{ik}')$\;
    $q\gets q+\Delta$\Comment{Add small noises}\;
    $z\gets q\le r$\Comment{Calculate is-zero Booleans}\;
    $d_{ik}''\gets\mathcal{Q}^{-1}(q)^T\cdot z$\Comment{Map values by quantile in original values}\;
    $d_{ik}''\gets d_{ik}''\times S\div s$\Comment{Scale degrees}\;
    $d_{ik}'\gets d_{ik}''\times w+d_{ik}'\times (1 - w)$\Comment{Weighted sum}\;
    $d_{ik}'\gets$ clipped based on $d_{\min},d_{\max}$\;
    $S_{\min},S_{\max}\gets S\times(1-\varepsilon),S\times(1+\varepsilon)$\;
    $S_{\min}',S_{\max}'\gets\sum\lfloor d_{ik}'\rfloor,\sum\lceil d_{ik}'\rceil$\Comment{Possible range by rounding}\;
    \If{$S_{\max}'<S_{\min}$}{
    $d_{ik}'\gets \lceil d_{ik}'\rceil$\Comment{Get ceils}\;
    $\delta\gets\lceil S_{\min}-S_{\max}'\rceil$\Comment{Calculate difference}\;
    $d_{ik}'\gets$ randomly increase degrees (capped at $d_{\max}$)
    }
    \ElseIf{$S_{\min}'>S_{\max}$}{
    $d_{ik}'\gets\lfloor d_{ik}'\rfloor$\Comment{Get floors}\;
    $\delta\gets\lfloor S_{\min}'-S_{\max}\rfloor$\Comment{Calculate difference}\;
    $d_{ik}'\gets$ randomly decrease degrees (capped at $d_{\min}$)
    }
    \Else{
    $t\gets$ rounding threshold s.t. $\sum$\texttt{round}$(d_{ik},t)\in[S_{\min},S_{\max}]$\;
    \Comment{Rounding threshold is 0.5 for classical rounding, 0 for floors, }\;
    \Comment{and 1 for ceilings, and obtained by binary search}\;
    $d_{ik}'\gets$\texttt{round}$(d_{ik},t)$
    }
    }
\end{algorithm}

%% file: algos/deg-edit.tex
\begin{algorithm}[t]
    \small
    \caption{Editing Degrees for Matching Constraints on $\phi_{ik}$}\label{alg:deg-edit}
    \DontPrintSemicolon
    \KwIn{\;
    $\tau$\Comment{Generated NULL indicators from ``n'' ($\tau_{ik}$)}\;
    $d$\Comment{Generated degrees ($d_{ik}$ after ``g'' with sum equal to $n_i-\|\tau\|$}\;
    $\mathcal{J}=[\mathcal{J}_j]_{j=1}^{n_j}$\;\Comment{Valid pools of parent row indices by valid-pair constraints}\;
    \KwOut{$d^*$\Comment{Edited $d$ allowing satisfaction of constraints}}
    $\mathcal{G}=[\mathcal{G}_j]_{j=1}^N\gets$ detected connected components in $\mathcal{J}$\;
    \Comment{Parent row indices as vertices, and all elements in $\mathcal{J}_j$ are\\\quad\quad\quad\quad mutually connected for all $j$, and $N$ is the number of\\\quad\quad\quad\quad components}\;
    \For{$x\in\{1,\dots,n_{F_{ik}}\}\setminus\bigcup_{j=1}^{n_j}\mathcal{J}_j$}{
    insert $\{x\}$ as an additional component in $\mathcal{G}$\;
    \Comment{Parent rows not in any pool}
    }
    $\boldsymbol{\delta}^*=[\delta_j^*]_{j=1}^N\gets[\sum_{x\in\mathcal{G}_j}d_x]_{j=1}^N$\;
    \Comment{Generated sums of degrees from $d$ in each $\mathcal{G}_j$}\;
    $\boldsymbol{\delta}^\dagger=[\delta_j^\dagger]_{j=1}^N\gets\left[\left|\{\mathcal{J}_{j'}\mid\mathcal{J}_{j'}\cap\mathcal{G}_j\ne\emptyset\text{ and } \tau_{j'}=0\}\right|\right]_{j=1}^N$\;
    \Comment{Minimum sum degrees required in each $\mathcal{G}_j$}\;
    $d^*\gets d$\Comment{Initialize modified degrees}\;
    \For{$\mathcal{G}_j\in\mathcal{G}$}{
        $\mathcal{X}^*\gets\mathtt{sample}([1,\dots,N], \mathtt{max\_cnts=}\boldsymbol{\delta}^\dagger-\boldsymbol{\delta}^*,\mathtt{size}=\delta_j^*-\delta_j^\dagger)$\;
        \Comment{Sample components with spare degrees}\;
        \For{$x\in\mathcal{X}^*$}{
            $\mathcal{Y}\gets\{y\mid y\in\mathcal{G}_x\text{ and }d^*_x>0\}$\;
            \Comment{Parent row indices with spare degrees in the component}\;
            $y\gets\mathtt{sample}(\mathcal{Y})$\;\Comment{Sampled parent row index to swap degree out}\;
            $z\gets\mathtt{sample}(\mathcal{G}_j)$\;\Comment{Sampled parent row index to swap degree in}\;
            $d^*_y,d^*_z\gets d^*_y-1,d_z^*+1$\;
            $\delta_x^*,\delta^*_j\gets\delta_x^*-1,\delta_j^*+1$\Comment{Update modifed degrees}\;
        }
    }
    }
\end{algorithm}

%% file: tables/football-fks.tex
\begin{tabular}{llp{0.5\linewidth}}
    \toprule
    Abbr. & Reference & Remark \\
    \midrule
    GT & TS.GID$\rightarrow$G.GID & part of composite PK \\
    T & TS.TID$\rightarrow$T.TID & part of composite PK \\
    G & G.GID$\rightarrow$GS.GID & FK as PK, total participation \\
    HT & G.(GID,HTID)$\rightarrow$TS.(GID,TID) & part of composite PK, composite FK, overlapping with another FK with non-self-match \\
    AT & G.(GID,ATID)$\rightarrow$TS.(GID,TID) & part of composite PK, composite FK, overlapping with another FK with non-self-match \\
    GA & A.GID$\rightarrow$G.GID & part of composite PK \\
    P & A.PID$\rightarrow$P.PID & part of composite PK \\
    SS & S.(GID,SID)$\rightarrow$A.(GID,PID) & composite FK, overlapping with another FK with inequality\\
    SA & S.(GID,AID)$\rightarrow$A.(GID,PID) & composite FK, overlapping with another FK with inequality, NULL-able \\
    \bottomrule
\end{tabular}

%% file: tables/sdv-football.tex
\begin{tabular}{lrr}
\hline
\textbf{Table Name} & \textbf{\# Columns in Metadata} & \textbf{Est. \# Columns} \\
\hline
teams        & 1  & \textcolor{red}{5{,}309{,}956{,}230{,}854{,}316} \\
players      & 1  & 10{,}149 \\
games        & 0  & 10{,}148 \\
gamestats    & 31 & \textcolor{red}{5{,}309{,}956{,}230{,}854{,}346} \\
appearances  & 17 & 139 \\
teamstats    & 14 & \textcolor{red}{103{,}052{,}956} \\
shots        & 8  & 8 \\
\hline
\end{tabular}


%% file: tables/timing.tex
\begin{tabular}{lcccc}
\toprule
{} &          IND~\cite{sdv-new} &        RCT~\cite{rctgan} &         CLD~\cite{clavaddpm} &          IRG \\
\midrule
train    &  3709 &  22661 &  39938 &  69699 \\
generate &    21 &    308 &    629 &  93970\\
\bottomrule
\end{tabular}

%% file: tables/table-resources.tex
\begin{tabular}{l|rrrrrrr|rrrrrrr|rrrrrrr}
\toprule
{} & \multicolumn{7}{c|}{CPU Mem. (MB)} & \multicolumn{7}{c|}{GPU Mem. (MB)} & \multicolumn{7}{c}{Time (s)} \\
{} &         \multicolumn{1}{c}{T} & \multicolumn{1}{c}{P} & \multicolumn{1}{c}{GS} & \multicolumn{1}{c}{TS} &   \multicolumn{1}{c}{G} & \multicolumn{1}{c}{A} &     \multicolumn{1}{c|}{S} &         \multicolumn{1}{c}{T} & \multicolumn{1}{c}{P} & \multicolumn{1}{c}{GS} & \multicolumn{1}{c}{TS} &    \multicolumn{1}{c}{G} & \multicolumn{1}{c}{A} &    \multicolumn{1}{c|}{S} &    \multicolumn{1}{c}{T} & \multicolumn{1}{c}{P} & \multicolumn{1}{c}{GS} & \multicolumn{1}{c}{TS} &   \multicolumn{1}{c}{G} & \multicolumn{1}{c}{A} &    \multicolumn{1}{c}{S} \\
\midrule
Fit data transformer               &        641.39 &  752.05 &    777.06 &    784.12 &  828.57 &     1096.02 &   2325.51 &          0.00 &    0.00 &      0.00 &      0.00 &     0.00 &        0.00 &     0.00 &     0.07 &    0.00 &      0.00 &      0.04 &    0.00 &        0.25 &    20.11 \\
Transform table with FK            &             - &       - &         - &    785.27 &  828.57 &     1105.32 &   2361.52 &             - &       - &         - &      0.00 &     0.00 &        0.00 &     0.00 &        - &       - &         - &      0.04 &    0.04 &        0.25 &    21.33 \\
Extend table                       &             - &       - &         - &    786.27 &  831.32 &     2093.90 &  19105.55 &             - &       - &         - &      0.00 &     0.00 &        0.00 &     0.00 &        - &       - &         - &      0.06 &    0.31 &        7.61 &    69.69 \\
Collect context for "a" and "d"    &             - &       - &         - &    828.57 &  932.85 &     1591.73 &  21814.97 &             - &       - &         - &      0.00 &     0.00 &        0.00 &     0.00 &        - &       - &         - &      0.01 &    0.50 &        0.29 &    11.19 \\
Train aggregated information       &             - &       - &         - &   2754.28 & 2971.97 &     3384.06 &  30514.32 &             - &       - &         - &   2422.96 &  3347.85 &     2661.92 & 20858.68 &        - &       - &         - &    259.15 &  116.96 &     2103.34 & 15233.71 \\
Train standalone/actual generation &       2566.82 & 2543.77 &   3017.45 &   3510.61 & 3093.39 &    15794.42 &  32773.22 &        848.37 &  854.60 &   1642.78 &  23484.86 &  7762.84 &    23003.20 & 23501.77 &   124.31 &  113.60 &   1262.72 &  13500.85 & 4166.06 &    25772.16 &  6821.29 \\
\midrule
Generate standalone/actual         &        810.86 & 1524.57 &   1350.85 &   2955.88 & 4179.91 &     6093.30 & 104512.43 &        203.43 &  203.43 &   1212.66 &  23594.33 & 23475.87 &    23460.25 & 23461.98 &     1.00 &    6.63 &     91.27 &   1604.93 &  893.84 &    40667.87 & 32094.75 \\
Generate aggregated information    &             - &       - &         - &   1676.80 & 3799.61 &     2991.27 &  67060.50 &             - &       - &         - &   1880.67 & 23235.69 &    13876.96 & 23435.98 &        - &       - &         - &      2.68 &   52.40 &       55.41 & 13392.94 \\
Prepare context for next table     &       1250.80 & 1283.84 &   1284.54 &   2833.76 & 2992.24 &    46778.00 &  66218.67 &          8.12 &    8.12 &      8.12 &      8.12 &     8.12 &        8.12 &     8.12 &     0.00 &    0.01 &      0.17 &      0.57 &    9.42 &       26.19 &   225.29 \\
\bottomrule
\end{tabular}

%% file: tables/fk-resources.tex
\begin{tabular}{l|rrrrrrrrr|rrrrrrrrr|rrrrrrrrr}
\toprule
{} & \multicolumn{9}{c|}{CPU Mem. (MB)} & \multicolumn{9}{c|}{GPU Mem. (MB)} & \multicolumn{9}{c}{Time (s)} \\
{} &             \multicolumn{1}{c}{T} &       \multicolumn{1}{c}{GT} &       \multicolumn{1}{c}{G} &      \multicolumn{1}{c}{HT} &      \multicolumn{1}{c}{AT} &       \multicolumn{1}{c}{P} &        \multicolumn{1}{c}{GA} &       \multicolumn{1}{c}{SS} &       \multicolumn{1}{c|}{S} &             \multicolumn{1}{c}{T} &     \multicolumn{1}{c}{GT} &     \multicolumn{1}{c}{G} &     \multicolumn{1}{c}{HT} &     \multicolumn{1}{c}{AT} &       \multicolumn{1}{c}{P} &      \multicolumn{1}{c}{GA} &    \multicolumn{1}{c}{SS} &     \multicolumn{1}{c|}{S} &        \multicolumn{1}{c}{T} &     \multicolumn{1}{c}{GT} &    \multicolumn{1}{c}{G} &   \multicolumn{1}{c}{HT} &   \multicolumn{1}{c}{AT} &    \multicolumn{1}{c}{P} &      \multicolumn{1}{c}{GA} &    \multicolumn{1}{c}{SS} &      \multicolumn{1}{c}{SA} \\
\midrule
Obtain parent extended till   &        828.57 &   828.57 &  932.85 &  932.85 &  932.25 & 1591.73 &   2076.18 & 15840.50 & 35012.08 &          0.00 &   0.00 &  0.00 &   0.00 &   0.00 &    0.00 &    0.00 &  0.00 &   0.00 &     0.00 &   0.02 & 0.09 & 0.08 & 0.17 & 0.01 &    2.58 &  9.95 &    9.85 \\
Fit degree regressor          &       2579.07 &  2582.33 & 2862.70 & 2879.45 & 2883.95 & 2980.28 &   3493.60 & 22718.01 & 27918.49 &        524.26 & 524.26 & 16.25 &  16.25 &  16.25 & 1234.92 & 1234.92 & 16.25 &  16.25 &     0.01 &   0.02 & 0.06 & 0.07 & 0.06 & 0.17 &    2.60 & 26.56 &   26.73 \\
Prepare NULL indicator inputs &             - &        - &       - &       - &       - &       - &         - &        - & 29505.34 &             - &      - &     - &      - &      - &       - &       - &     - &   0.00 &        - &      - &    - &    - &    - &    - &       - &     - &    0.60 \\
Fit NULL indicator classifier &             - &        - &       - &       - &       - &       - &         - &        - & 17600.11 &             - &      - &     - &      - &      - &       - &       - &     - &  16.25 &        - &      - &    - &    - &    - &    - &       - &     - &   12.96 \\
\midrule
Generate degrees              &       1301.25 &  2827.90 & 2833.76 & 2984.41 & 2981.41 & 2991.27 &   4441.66 & 42769.75 & 65857.30 &          8.12 &   8.12 &  8.12 &   8.12 &   8.12 &    8.12 &    8.12 &  8.12 &   8.12 &     0.01 &   0.02 & 0.05 & 0.12 & 0.11 & 0.12 &    0.68 &  8.18 &   18.54 \\
Generate NULL indicator       &             - &  - &       - & - & - &       - &   - &        - & 69795.49 &             - &   - &     - &   - &  -&       - &   - &     - &   8.12 &        - &   - &    - & - & - &    - &   - &     - &   19.54 \\
Match FK for                     &             - & 16390.93 &       - & 8877.29 & 8732.09 &       - & 110179.20 &        - & 76875.70 &             - & 500.83 &     - & 997.88 & 997.88 &       - &  773.50 &     - & 714.55 &        - & 267.34 &    - & 4.30 & 4.91 &    - & 3102.51 &     - & 1397.82 \\
\bottomrule
\end{tabular}

%% file: tables/football-pk.tex
\begin{tabular}{lccc>{\columncolor[gray]{0.9}}c}
\toprule
               &             IND~\cite{sdv-new} &                  RCT~\cite{rctgan} &                   CLD~\cite{clavaddpm} &             IRG \\
\midrule
 A &  0.000 &  \textcolor{red}{0.004} &  \textcolor{red}{0.028} &  0.000 \\
 TS &  0.000 &  \textcolor{red}{0.003} &  \textcolor{red}{0.003} &  0.000 \\
\midrule
     Avg. &  0.000 &  \textcolor{red}{0.004} &  \textcolor{red}{0.015} &  0.000 \\
     \# Vio. &  0/2 &  \textcolor{red}{2/2} &  \textcolor{red}{2/2} &  0/2 \\
\bottomrule
\end{tabular}

%% file: tables/football-fk-acc.tex
\begin{tabular}{lccc>{\columncolor[gray]{0.9}}c}
\toprule
          &             IND~\cite{sdv-new} &                  RCT~\cite{rctgan} &                   CLD~\cite{clavaddpm} &             IRG \\
\midrule
G &             0.000 &                \textcolor{red}{0.039} &                     0.000 &             0.000 \\
     HT &             \textcolor{red}{-} &                \textcolor{red}{0.985} &                \textcolor{red}{0.977} &             0.000 \\
     AT &             \textcolor{red}{-} &                \textcolor{red}{0.988} &                \textcolor{red}{0.977} &             0.000 \\
     P &        \textcolor{red}{0.979} &                     0.000 &                     0.000 &             0.000 \\
     GA &        \textcolor{red}{1.000} &                     0.000 &                \textcolor{red}{0.064} &             0.000 \\
     T &             \textcolor{red}{1.000} &                     0.000 &                     0.000 &             0.000 \\
     GT &             \textcolor{red}{1.000} &                     0.000 &                     0.000 &             0.000 \\
     SS &             0.000 &                     0.000 &                     0.000 &             0.000 \\
     SA &             \textcolor{red}{-} &                \textcolor{red}{0.991} &                \textcolor{red}{0.871} &             0.000 \\
     \midrule
      Avg. &        \textcolor{red}{0.663} &                \textcolor{red}{0.334} &                \textcolor{red}{0.321} &             0.000 \\
     \# Vio. &             \textcolor{red}{7/9} &                     \textcolor{red}{4/9} &                     \textcolor{red}{4/9} &             0/9 \\
\bottomrule
\end{tabular}

%% file: tables/ineq.tex
\begin{tabular}{lccc>{\columncolor[gray]{0.9}}c}
\toprule
            &             IND~\cite{sdv-new} &                  RCT~\cite{rctgan} &                   CLD~\cite{clavaddpm} &             IRG \\
\midrule
 G &  0.000 &  \textcolor{red}{0.007} &  \textcolor{red}{0.071} &  0.000 \\
     S &  0.000 &  \textcolor{red}{0.000} &  \textcolor{red}{0.057} &  0.000 \\
\midrule
     Avg. &  0.000 &  \textcolor{red}{0.004} &  \textcolor{red}{0.064} &  0.000 \\
     \# Vio. &  0/2 &  \textcolor{red}{2/2} &  \textcolor{red}{2/2} &  0/2\\
\bottomrule
\end{tabular}

%% file: tables/deg-dist.tex
\begin{tabular}{lccc>{\columncolor[gray]{0.9}}c}
\toprule
FK &             IND~\cite{sdv-new} & RCT~\cite{rctgan} &           CLD~\cite{clavaddpm}&    IRG \\
\midrule
G                   &           \textbf{0.000} &  \textbf{0.000} &  0.834 &  \textbf{0.000} \\
HT &  0.500 &  0.492 &           0.489 &  \textbf{0.000} \\
AT &  0.500 &  0.494 &           0.489 &  \textbf{0.000} \\
P           &           0.773 &  0.033 &  0.838 &  \textbf{0.001} \\
GA                 &  1.000 &  0.596 &  1.000 &  \textbf{0.068} \\
T                   &  1.000 &  0.466 &           0.836 &  \textbf{0.007} \\
GT               &  1.000 &  \textbf{0.000} &           0.971 &  \textbf{0.000} \\
SS &  0.473 &  \textbf{0.051} &           0.104 &  0.054 \\
SA &  0.395 &  0.388 &           0.302 &  \textbf{0.193} \\
\midrule
Avg.                                               &           0.627 &  0.280 &  0.651 &  \textbf{0.036} \\
\bottomrule
\end{tabular}

%% file: tables/table-sizes.tex
\begin{tabular}{lccc>{\columncolor[gray]{0.9}}c}
\toprule
{} &             IND~\cite{sdv-new} &          RCT~\cite{rctgan} &                    CLD~\cite{clavaddpm}&             IRG \\
\midrule
T       &  \textbf{0.000} &  \textbf{0.000} &           \textbf{0.000} &  \textbf{0.000} \\
P     &  \textbf{0.000} &  \textbf{0.000} &           \textbf{0.000} &  \textbf{0.000} \\
GS   &  \textbf{0.000} &  \textbf{0.000} &           \textbf{0.000} &  \textbf{0.000} \\
G       &  \textbf{0.000} &           0.040 &  \textcolor{red}{-0.834} &  \textbf{0.000} \\
A &  \textbf{0.000} &          -0.078 &  \textcolor{red}{-0.994} &  \textbf{0.000} \\
TS   &  \textbf{0.000} &  \textbf{0.000} &  \textcolor{red}{-0.821} &  \textbf{0.000} \\
S       &  \textbf{0.000} &           0.139 &  \textcolor{red}{-0.993} &  \textbf{0.000} \\
\midrule
Avg.        &  \textbf{0.000} &           0.037 &   \textcolor{red}{0.520} &  \textbf{0.000} \\
\bottomrule
\end{tabular}

%% file: tables/olist-pk.tex
\begin{tabular}{lccc>{\columncolor[gray]{0.9}}c}
\toprule
{} &                     IND~\cite{sdv-new} &                  RCT~\cite{rctgan} &                   CLD~\cite{clavaddpm}&             IRG \\
\midrule
OI    &  \textcolor{red}{0.000} &  \textcolor{red}{0.085} &  \textcolor{red}{0.101} &  0.000 \\
OP &          0.000 &  \textcolor{red}{0.015} &  \textcolor{red}{0.037} &  0.000 \\
\midrule
Avg.           &  \textcolor{red}{0.000} &  \textcolor{red}{0.050} &  \textcolor{red}{0.069} &  0.000 \\
\# Vio.         &  \textcolor{red}{1/2} &  \textcolor{red}{2/2} &  \textcolor{red}{2/2} &  0/2 \\
\bottomrule
\end{tabular}

%% file: tables/smm-fk-acc.tex
\begin{tabular}{lcc>{\columncolor[gray]{0.9}}c}
    \toprule
     & IND~\cite{sdv-new} & CLD~\cite{clavaddpm} & IRG \\
    \midrule
    Avg. & \textcolor{red}{0.505} & \textcolor{red}{0.111} & 0.000 \\
    \# Vio. & \textcolor{red}{5/9} & \textcolor{red}{2/9} & 0/9 \\
    \bottomrule
\end{tabular}

%% file: tables/smm-pks.tex
\begin{tabular}{lcc>{\columncolor[gray]{0.9}}c}
\toprule
{} &                              IND~\cite{sdv-new} &                            CLD~\cite{clavaddpm} & IRG \\
\midrule
\texttt{plays}   &                   0.000 &           \textcolor{red}{0.000} & 0.000 \\
\texttt{clears}  &           \textcolor{red}{1.000} &  \textcolor{red}{0.005} & 0.000 \\
\texttt{likes}   &           \textcolor{red}{1.000} &  \textcolor{red}{0.004} & 0.000 \\
\texttt{records} &           \textcolor{red}{1.000} &  \textcolor{red}{0.004} & 0.000 \\
\midrule
Avg.    &           \textcolor{red}{0.750} &  \textcolor{red}{0.003} & 0.000 \\
\# Vio.  &  \textcolor{red}{3/4} &           \textcolor{red}{4/4} & 0/4 \\
\bottomrule
\end{tabular}

%% file: tables/abl-sum-deg.tex
\begin{tabular}{lccc}
\toprule
{} & Exp. sum & IRG sum & w/o sum ctrl. \\
\midrule
T & 25360 & 25360 & 25404 \\
GT & 25360 & 25360 & 25360 \\
G & 356513 & 356513 & 351296 \\
HT & 12680 & 12680 & 12680 \\
AT & 12680 & 12680 & 12345 \\
P & 356513 & 356513 & 303385 \\
SS & 324543 & 324543 & 255923 \\
SA & 232704 & 232704 & 87359 \\
\bottomrule
\end{tabular}

%% file: tables/ablation-component.tex
\begin{tabular}{lcccc}
    \toprule
    Model & Real & IRG & w/o seq. & small comp. \\
    \midrule
    DT & $0.504_{\pm 0.006}$ & $0.357_{\pm0.005}$ & $0.324_{\pm0.004}$ & $0.358_{\pm0.007}$\\
    RF & $0.596_{\pm0.001}$ & $0.310_{\pm0.004}$ & $0.344_{\pm0.009}$ & $0.286_{\pm0.006}$ \\
    XGB & $0.581_{\pm0.006}$ & $0.304_{\pm0.000}$ & $0.302_{\pm0.000}$ & $0.289_{\pm0.000}$\\
    LGBM & $0.597_{\pm0.009}$ & $0.320_{\pm0.000}$ & $0.307_{\pm0.000}$ & $0.278_{\pm0.003}$\\
    \midrule
    Avg. & 0.570 & 0.323 & 0.319 & 0.303\\
    \bottomrule
\end{tabular}